\documentclass[8.5pt,twoside,twocolumn]{article}
\usepackage[super,sort&compress,comma]{natbib} 
\usepackage{mhchem}
\usepackage{times,mathptmx}
\usepackage{sectsty}
\usepackage{balance} 
\usepackage{graphicx} %eps figures can be used instead
\usepackage{lastpage}
\usepackage[format=plain,justification=raggedright,singlelinecheck=false,font=small,labelfont=bf,labelsep=space]{caption} 
\usepackage{fancyhdr}
\usepackage{amssymb}
\usepackage{epsfig} % figuras no formato eps (Encapsulated PostScript)
\usepackage{subfigure} % figuras lado a lado
\usepackage[english]{babel}
\usepackage{amsmath}
\usepackage{textcomp}
\usepackage{color}
\usepackage{amsfonts}
\usepackage{hyperref}
\usepackage{float}
\usepackage{scalefnt}
\usepackage{braket}
\usepackage{ulem}

\newcommand{\onlinecite}[1]{\hspace{-1 ex} \nocite{#1}\citenum{#1}}

\begin{document}

\thispagestyle{plain}
\fancypagestyle{plain}{
%\fancyhead[L]{\includegraphics[height=8pt]{headers/LH}}
%\fancyhead[C]{\hspace{-1cm}\includegraphics[height=20pt]{headers/CH}}
%\fancyhead[R]{\includegraphics[height=10pt]{headers/RH}\vspace{-0.2cm}}
\renewcommand{\headrulewidth}{1pt}}
\renewcommand{\thefootnote}{\fnsymbol{footnote}}
\renewcommand\footnoterule{\vspace*{1pt}% 
\hrule width 3.4in height 0.4pt \vspace*{5pt}} 
\setcounter{secnumdepth}{5}

\makeatletter 
\def\subsubsection{\@startsection{subsubsection}{3}{10pt}{-1.25ex plus -1ex minus -.1ex}{0ex plus 0ex}{\normalsize\bf}} 
\def\paragraph{\@startsection{paragraph}{4}{10pt}{-1.25ex plus -1ex minus -.1ex}{0ex plus 0ex}{\normalsize\textit}} 
\renewcommand\@biblabel[1]{#1}
\renewcommand\@makefntext[1]% 
{\noindent\makebox[0pt][r]{\@thefnmark\,}#1}
\makeatother 
\renewcommand{\figurename}{\small{Fig.}~}
\sectionfont{\large}
\subsectionfont{\normalsize} 

%\fancyfoot{}
%\fancyfoot[LO,RE]{\vspace{-7pt}\includegraphics[height=9pt]{headers/LF}}
%\fancyfoot[CO]{\vspace{-7.2pt}\hspace{12.2cm}\includegraphics{headers/RF}}
%\%fancyfoot[CE]{\vspace{-7.5pt}\hspace{-13.5cm}\includegraphics{headers/RF}}
%\fancyfoot[RO]{\footnotesize{\sffamily{1--\pageref{LastPage} ~\textbar \hspace{2pt}\thepage}}}
%\fancyfoot[LE]{\footnotesize{\sffamily{\thepage~\textbar\hspace{3.45cm} 1--\pageref{LastPage}}}}
\fancyhead{}
\renewcommand{\headrulewidth}{1pt} 
\renewcommand{\footrulewidth}{1pt}
\setlength{\arrayrulewidth}{1pt}
\setlength{\columnsep}{6.5mm}
\setlength\bibsep{1pt}

\twocolumn[
  \begin{@twocolumnfalse}
\noindent\LARGE{\textbf{Stretching of BDT-gold molecular junctions: thiol or thiolate termination?}}
\vspace{0.6cm}

\noindent\large{\textbf{Amaury de Melo Souza,\textit{$^{a}$} Ivan Rungger,\textit{$^{a}$} 
Renato Borges Pontes,\textit{$^{b}$} Alexandre Reily Rocha,\textit{$^{c}$} Antonio Jose Roque da Silva,\textit{$^{d\ast}$} 
Udo Schwingenschloegl,\textit{$^{e}$} and Stefano Sanvito\textit{$^{a}$}}}\vspace{0.5cm}
%Please note that \ast indicates the corresponding author(s) but no footnote text is required. 

\noindent\textit{\small{\textbf{Received Xth XXXXXXXXXX 20XX, Accepted Xth XXXXXXXXX 20XX\newline
First published on the web Xth XXXXXXXXXX 200X}}}

\noindent \textbf{\small{DOI: 10.1039/b000000x}}
\vspace{0.6cm}
%Please do not change this text.

\noindent
\normalsize{
It is often assumed that the hydrogen atoms in the thiol groups of a benzene-1,4-dithiol dissociate
when Au-benzene-1,4-dithiol-Au junctions are formed. We demonstrate, by stability and transport
properties calculations, that this assumption can not be made. We show that the dissociative
adsorption of methanethiol and benzene-1,4-dithiol molecules on a flat Au(111) surface is
energetically unfavorable and that the activation barrier for this reaction is as high
as 1 eV. For the molecule in the junction, our results show, for all electrode geometries
studied, that the thiol junctions are energetically more stable than their thiolate counterparts.
Due to the fact that density functional theory (DFT) within the local density approximation (LDA)
underestimates the energy difference between the lowest unoccupied molecular orbital and the highest
occupied molecular orbital by several electron-volts, and that it does not capture the renormalization
of the energy levels due to the image charge effect, the conductance of the Au-benzene-1,4-dithiol-Au
junctions is overestimated. After taking into account corrections due to image charge effects by means
of constrained-DFT calculations and electrostatic classical models, we apply a scissor operator to
correct the DFT energy levels positions, and calculate the transport properties of the thiol and
thiolate molecular junctions as a function of the electrodes separation. For the thiol junctions,
we show that the conductance decreases as the electrodes separation increases, whereas the opposite
trend is found for the thiolate junctions. Both behaviors have been observed in experiments,
therefore pointing to the possible coexistence of both thiol and thiolate junctions. Moreover,
the corrected conductance values, for both thiol and thiolate, are up to two orders of magnitude
smaller than those calculated with DFT-LDA. This brings the theoretical results in quantitatively good agreement with experimental data.} 
\vspace{0.5cm}
\end{@twocolumnfalse}
]

\section{Introduction}\label{bdt-intro}

\footnotetext{\textit{$^{a}$~School of Physics; AMBER and CRANN, Trinity College Dublin, College Green D2, Ireland, E-mail: souzaa@tcd.ie}}
\footnotetext{\textit{$^{b}$~Instituto de F\'isica, Universidade Federal de Goi\'as, Campus Samambaia 74001-970 , Goi\^ania-GO, Brasil}}
\footnotetext{\textit{$^{c}$~Instituto de F\'isica Te\'orica - Universidade Estadual Paulista, Barra-Funda 01140-070, S\~ao Paulo-SP, Brazil}}
\footnotetext{\textit{$^{d}$~Instituto de F\'isica, Universidade de S\~ao Paulo, Rua do Mat\~ao 05508-090, Cidade Universit\'aria, S\~ao Paulo-SP, Brazil}}
\footnotetext{\textit{$^{\ast}$~Laborat\'orio Nacional de Luz Sincroton LNLS, 13083-970 Campinas-SP, Brazil}}
\footnotetext{\textit{$^{e}$~PSE Division, KAUST, Thuwal 23955-6900, Saudi Arabia}}

%=====================================================
% The main text starts here!
A long standing problem in the area of molecular electronics has been the difficulty of finding quantitative agreement 
between theory and experiment in some cases. This makes it difficult to design and build functioning devices based on molecules.
More than a decade has passed since the pioneering experiment by Reed \textit{et al.},\cite{Reed1997} and yet the
well-known prototype molecular junction that consists of a benzene-1,4-dithiol molecule between two gold electrodes
is still not fully understood. Numerous experimental~\cite{Tsutsui2006,Tsutsui2009,Tian2010,Tao2004,Taniguchi2010,Baheti2008,Lortscher2007,Baheti2008,Kim2011}
and theoretical~\cite{Demir2012,Pontes2006,Pontes2011a,Toher2008,Strange2011a,French2013,French2013a} works have been reported,
with both experimental and theoretical results varying over a large range. 

In general, the possible experimental setups can be divided into two main categories: mechanically controlled break-junctions
(MCBJs)~\cite{Reed1997,Lortscher2007,Gonzalez2006,Kim2011, Taniguchi2010,Tian2006,Tian2010,Tsutsui2009,Tsutsui2006,Tsutsui2009a}
and scanning tunneling microscopy (STM) experiments.\cite{Quek2009, Arroyo2011, Bruot2012, Baheti2008,Fatemi2011,Kiguchi2010,Reddy2007,Vazquez2012,
Venkataraman2006,Wold2002,Tao2004,Xu2003} In the former, a gold nano-contact is created by stretching a gold wire and,
just before rupture, a solution containing the target molecules is added to the system. Subsequently, the metallic contact
is further stretched until rupture and in some cases the molecules remain trapped between the Au-Au tips forming the molecular
junctions. In the second setup, the target molecules are deposited on a gold surface, and a STM tip is brought into contact to form
the junction. Due to the nature of the experiment, several different geometrical contacts
can be accessed during the stretching process of
the junction, which leads to a statistical character of the experimental analysis. 
In fact, in a single experiment, a broad range of values of conductance, $G$,   is observed, and possibly even
very different average $G$ values between experiments.\cite{Tsutsui2006,Kim2011,Pontes2011a} Yet, recent independent
measurements~\cite{Tao2004,Bruot2012,Kiguchi2010,Tsutsui2009a} agree on an average value of $G$ of about 0.01$G_0$,
where $G_0=2e^2/h$  is the quantum conductance ($e$ is the electron charge and $h$ is the Planck's constant).

From the theoretical point of view, the quantitative description of such molecular junctions is challenging for two main reasons.
Firstly, realistic electrode configurations and many arrangements should be considered in the calculations,
which becomes prohibitive within a fully \textit{ab initio} approach. More recently, French \textit{et al.}~\cite{French2013, French2013a}
have applied a sophisticated method that combines Monte Carlo simulations and classical molecular dynamics to simulate the junction stretching process,
allowing the sampling of hundreds of contact geometries between the molecule and the electrodes. In addition, it is generally assumed in
the literature~\cite{Tomfohr2004,Stokbro2003,Maksymovych2008,Andreoni2000,French2013, French2013a,Pu2010,Pontes2006,Pontes2011a,Strange2011,
Strange2011a,Toher2008,Gronbeck2000} that when the molecule attaches to the gold electrodes,
the hydrogen atoms linked to the thiol groups are dissociated to form a thiolate-Au bond.
However, recent DFT calculations on the details of the adsorption of the benzene-1,4-dithiol on gold have
been reported.\cite{Nara2004, Ning2010} They find that the thiol-Au structure is energetically
more stable than its thiolate-Au counterparts when the molecule binds to either a perfect flat
surface~\cite{Nara2004} or to an adatom structure.\cite{Ning2010} 

Usually transport calculations rely on the Kohn-Sham (KS) eigenvalues to evaluate $G$, even though these eigenvalues can not be
rigorously interpreted as quasi-particle energy levels. The only exception is for the HOMO level, which is equal to the negative
of the ionization potential.\cite{Perdew1983,Parr1979,J.F.Janak1978} It has been demonstrated
experimentally~\cite{Search1997,Repp2005,Lu2004,Greiner2011,Perrin2013} that the quasi-particle
energy gap, $E_\mathrm{QP}^\mathrm{gap}$,  of a molecule, defined as the difference between its
ionization potential, $I$, and electron affinity, $A$, shrinks with respect to that of the gas
phase by adsorbing the molecule on a polarizable substrate. Nevertheless, the electronic structure
theories usually used for such calculations can only partly account for this renormalization of the
molecular energy levels when the junction is formed. It is well-known that DFT, within the standard local and semi-local
approximations to the exchange-correlation (XC) energy, does not include non-local correlation effects,
such as the dynamical response of the electron system to adding electrons or holes to the molecule.
This limits its ability to predict the energy level alignment, when compared to experiments,
which often leads to overestimated values for $G$.\cite{Kronik2008, Flores2009} 

A rigorous way to include such non-local correlation effects is by using many-body perturbation theory,
such as for example the GW approximation constructed on top of DFT.\cite{Search1973,Hybertsen1986,Onida2002}
In the last few years, this approach has been used for evaluating level alignments,\cite{Neaton2006,Garcia-Lastra2011,Garcia-Lastra2009,
Tamblyn2011,Rignanese2001,Strange2012} in general with good success.
The drawback of the GW scheme lays on the fact that it is highly computationally demanding,
which limits the system size that can be tackled. This is particularly critical in the case
of molecular junctions, where the system can be considerably large due to the presence of the
metal electrodes. Therefore, different alternative approaches and corrections have been proposed
to improve the description of the energy level alignment, for instance, corrections for self-interaction
(SI) errors,\cite{Toher2008, Pemmaraju2007} scissor operator (SCO)
schemes~\cite{Quek2007,Strange2011,Neaton2006,Flores2009,Garcia-Suarez2011} and constrained-DFT (CDFT).\cite{Souza2013}

In the present work we investigate, by means of total energy DFT and quantum transport calculations,
the stability and conductivity of thiol and thiolate molecular junctions. We compare the results for the
two systems and we relate them to experimental data. The paper is divided as follows. In Sec. \ref{methods}
we first give an overview of the methodology used.
In Sec. \ref{BDT-adsorption} we present a systematic study of the adsorption process of
two thiol-terminated molecules, namely, methanethiol and benzene-1,4-dithiol on Au(111) flat surface.
For the latter, we also compare the stability of the thiol and thiolate systems when the junction is
formed for several contact geometries.\cite{Pontes2006, Pontes2011a, Toher2008, French2013, French2013a}
In Sec. \ref{level-alignment} we discuss the energy level alignment, and present three methods used to
correct the DFT-LDA molecular energy levels, namely CDFT and SCO. Based on these results in Sec. \ref{Transport}
we finally discuss the transport properties and present the dependence of $G$ on the electrodes separation ($L$)
for flat-flat contact geometries, for both the thiol and thiolate junctions. 
\section{METHODS}\label{methods}
\subsection{Calculation details}

All the calculations presented in this paper are based on DFT as implemented in the {\sc siesta} package.\cite{Soler2002}
For some calculations, we also use the plane-wave code {\sc vasp}~\cite{Kresse1996}
in order to compare with our results obtained with the localized basis set.
Unless stated otherwise, we use the following parameters throughout this work.
For total energy and relaxation calculations we use the generalized gradient approximation
as formulated by Perdew-Burke-Ernzenhof (GGA-PBE) to the XC energy.\cite{Perdew1996} The basis set for {\sc siesta}
is the double-$\zeta$ polarized for carbon, sulfur and hydrogen and
a double-$\zeta$ for the 5$d$6$s$6$p$ orbitals of Au atoms. 
We take into account corrections for the basis set superposition error (BSSE).
The mesh cutoff
is 300 Ry, four $k$-points are used for the Brillouin zone sampling in the perpendicular
direction to the transport and norm-conserving pseudopotentials according to the Troullier-Martins
procedure to describe the core electrons.\cite{Troullier1991} In the case of {\sc vasp} calculations,
we use a cut-off energy of 450 Ry to expand the wave functions and the projector augmented-wave method to
treat the core electrons.\cite{Rostgaard2009} All the junctions are fully relaxed until all the forces
are smaller than 0.02 eV/\AA.  All the quantum transport calculations presented are performed with
the {\sc smeagol} code,\cite{Rocha2006,Rungger2008} which uses the non-equilibrium Green's function (NEGF) formalism. Here,
the XC energy is treated within the LDA approximation.

\subsection{Self-interaction correction}\label{asic_methods}
One of the main deficiencies of local and semi-local DFT functionals when treating
organic/inorganic interfaces is the SI error. This spurious interaction of an electron with the Hartree 
and XC potentials generated by itself leads to an over-delocalization of the electronic charge density.
Consequently, the occupied KS eigenstates of molecules are pushed to higher energies.
Moreover, the unoccupied states are found too low in energy due to the lack of the
derivative discontinuity in the XC potential.\cite{Parr1979} These two limitations
lead to a substantial underestimation of the energy gap of various systems. 
In order to deal with the problem of SI, we apply the atomic self-interaction correction
(ASIC) method,\cite{Toher2008,Filippetti2011,Toher2005,Pemmaraju2007} which has been shown
to improve the position of the highest occupied molecular orbital (HOMO) of molecules
when compared to their gas phase $I$. It has also been shown to improve the energy level
alignment when the junction is formed, leading to values of $G$ in better agreement
with experiments.\cite{Pontes2011a,Toher2008, French2013, French2013a}
The method, however, shows some limitations. The correction applied by
ASIC depends on the atomic orbital occupation, not the molecular orbital occupation.
Therefore, if different molecular orbitals are composed of a linear combination of a similar
set of atomic orbitals, ASIC will shift their energy eigenvalues by a similar amount.
For example, if empty states share the same character as the occupied states,
as it is usually the case for small molecules, the energy of these states will be spuriously shifted
to lower energies. In order to apply ASIC, a scaling parameter, $\alpha$, to the atomic-like
occupations needs to be specified, where for $\alpha=1$ the full correction is applied,
while for $\alpha=0$ no correction is applied. The value of $\alpha$ is related to the screening provided by
the chemical environment.\cite{Pemmaraju2007} For metals, where the SI is negligible, we therefore use
$\alpha$=0, whereas for the molecules, where SI is more pronounced, we use $\alpha$=1. 
\subsection{Constrained Density Functional Theory}\label{cdft_methods}

In the present work we apply the CDFT method, described in Ref. [\onlinecite{Souza2013}],
to calculate the charge-transfer energy between the molecule and the metallic substrate.
 This corresponds to the position of the frontier energy levels, i.e., the HOMO and the
 lowest unoccupied molecular orbital (LUMO), with respect to the metal Fermi energy, $E_\mathrm{F}$.
 For a given distance, $d$, from the center of the molecule to the surface, the procedure is as follows:
 first, a conventional DFT calculation is performed, where no constraint is applied. This yields the total
 energy of the combined system, $E(\mathrm{mol}/\mathrm{sub};d)$, and the amount of charge present on each
 fragment (one fragment being the molecule and the other fragment the metal surface).
 In a second step, a CDFT calculation is performed.
Since we are interested in accessing the position of the frontier energy levels with respect to the
metal $E_\mathrm{F}$, we consider two types of constraints. In the first one, a full electron is
removed from the molecule and added to the substrate, and the total energy of this charge-transfer
state, $E(\mathrm{mol}^{+}/\mathrm{sub}^{-};d)$, is obtained. Hence, the energy to transfer one
electron from the molecule to the substrate is given by
\begin{equation}
E_\mathrm{CT}^{+}(d)=E(\mathrm{mol}^{+}/\mathrm{sub}^{-};d)-E(\mathrm{mol}/\mathrm{sub};d)\:.
\label{e-homo}
\end{equation} 
In the second case we evaluate the energy when one full electron is removed from the substrate and added
to the molecule, $E(\mathrm{mol}^{-}/\mathrm{sub}^{+};d)$. The charge-transfer energy to add one electron to the molecule is then given by 
\begin{equation}
E_{\mathrm{CT}}^{-}(d)=E(\mathrm{mol}/\mathrm{sub};d)-E(\mathrm{mol}^{-}/\mathrm{sub}^{+};d)\:.
\label{e-lumo}
\end{equation}
We can relate the charge-transfer energies to the frontier energy levels
by offsetting them with the metal work function ($W_\mathrm{F}$), so that $E_{\mathrm{HOMO}}(d)\simeq -[E^+_{\mathrm{CT}}(d)+W_{\mathrm{F}}]$ 
and $E_{\mathrm{LUMO}}(d)\simeq -[E^-_{\mathrm{CT}}(d)+W_{\mathrm{F}}]$ correspond to the HOMO and LUMO energies, respectively. 
Note that if the metal substrate is semi-infinite in size, then these relations are exact, since by definition 
the energy required to remove an electron from the metal and that gained by adding it are equal to $W_\mathrm{F}$.
However, in a practical calculation a finite size slab is used, and therefore, 
the relations are only approximately valid due to the inaccuracies in the calculated $W_\mathrm{F}$ for finite systems.
$W_\mathrm{F}$ is calculated by performing a simulation for the metal slab 
and by taking the difference between the vacuum potential and the $E_\mathrm{F}$ of the slab.

We can then compare the CDFT results for the renormalization of the energy levels due to the 
image charge effect with two simplified classical electrostatic models.
In the first one we consider the electrostatic energy of a point charge
interacting with a single surface,\cite{Lang1973} given by
\begin{equation}
V(d)=-\frac{q^2}{4(d-d_0)}\:.
\label{cm1s}
\end{equation}
In the second model, the point charge is interacting with two infinite flat surfaces,\cite{Quek2007,Garcia-Suarez2011}
which gives the following interaction energy
\begin{equation}
 U(d)=-\frac{q^2}{2(d-d_0)}\mathrm{ln2}\label{cm2s}.
\end{equation}
In both equations, $q$ is a point charge located at the center of the molecule,
and $d=L/2$ for the case of two surfaces, where $L$ is the distance between the two surfaces;
$d_0$ is the height of the image charge plane with respect to the surface atomic layer,
so that $d_0$ can be interpreted as the center of gravity 
of the screening charge density formed on the metal surface,
which in general depends on $d$. Instead of treating $d_0$ as a free parameter,
as usually done in the literature,\cite{Quek2007,Garcia-Suarez2011}
our CDFT approach allows us to calculate it from first principles.\cite{Souza2013} Hence the classical models
shown in Eq. (\ref{cm1s}) and Eq. (\ref{cm2s}) are effectively parameter-free when based on the CDFT value for $d_0$. 
\subsection{Scissor Operator method}\label{sco_methods}

Since we obtain the energies of the HOMO and LUMO from CDFT total energies,
we can shift the DFT eigenvalues to lie at these energies by means 
of a SCO~\cite{Ferretti2005,Quek2011,Garcia-Suarez2011,Quek2007,Mowbray2008,Abad2008}
method. This has been shown to improve $G$ when compared to experimental data.\cite{Quek2007}
For the particular case of a single molecule attached to the electrodes,
first a projection of the full KS-Hamiltonian matrix and of the overlap matrix
is carried out onto the atomic orbitals associated with the molecule subspace,
which we denote as $H_\mathrm{mol}^0$ and $S_\mathrm{mol}^0$ (the remaining part of $\hat{H}$ describes the electrodes). By
solving the corresponding eigenvalue problem,
$H_\mathrm{mol}^0 \psi=\epsilon S_\mathrm{mol}^0\psi$, for this subblock we
obtain the eigenvalues, $\{\epsilon_{n}\}_{n=1,...,M}$,
and eigenvectors, $\{\psi_{n}\}_{n=1,...,M}$,
where $M$ is the number of atomic orbitals on the molecule. Subsequently, the corrections are applied to the eigenvalues,
where all the occupied levels are shifted rigidly by the constant $\Sigma_\mathrm{o}$
while the unoccupied levels are shifted rigidly by the constant $\Sigma_\mathrm{u}$.
We note that in principle each state can be shifted by a different amount.
Using the shifted eigenvalues we can construct a transformed molecular Hamiltonian matrix,
$H_\mathrm{mol}^\mathrm{SCO}$, given by
\begin{equation}
 H_\mathrm{mol}^\mathrm{SCO}=H_\mathrm{mol}^0 + \Sigma_\mathrm{o}\sum_{i_\mathrm{o}=1,n_\mathrm{o}}\psi_{i_\mathrm{o}}\psi_{i_\mathrm{o}}^{\dagger}+\Sigma_\mathrm{u}\sum_{i_\mathrm{u}=1,n_\mathrm{u}}\psi_{i_\mathrm{u}}\psi_{i_\mathrm{u}}^{\dagger},\label{Eq.sco_method}
\end{equation}
where the first sum runs over the $n_\mathrm{o}$ occupied orbitals, and the second one runs over the $n_\mathrm{u}$ empty states.
In the full Hamiltonian matrix we then replace the subblock $H_\mathrm{mol}^0$ with
$H_\mathrm{mol}^\mathrm{SCO}$.\cite{Quek2011,Garcia-Suarez2011,Quek2007,Mowbray2008,Abad2008}
The SCO procedure can be applied self-consistently, although in this work we apply it non-selfconsistently to the converged DFT Hamiltonian.

The correction applied to the frontier energy levels of a molecule in a junction has two contributions.
First we need to correct for the fact that the gas-phase LDA HOMO-LUMO gap ($E_\mathrm{LDA}^\mathrm{gap}$) is
too small when compared to the difference between $I$ and $A$, where 
$I=E^{(N-1)}-E^{(N)}$ and $A= E^{(N)}-E^{(N+1)}$ ($E^{(N)}$ is the
ground state total energy for a system with $N$ electrons).
Secondly, the renormalization of the energy levels, when the molecule is brought close to metal surfaces needs to be added
to the gas-phase HOMO and LUMO levels. Although CDFT in principle allows us to assess the renormalization
of the energy levels in the junction, to reduce the computational costs we calculate the charge-transfer
energies with one single surface. Since in transport calculations there are two surfaces,
we then use the corresponding classical model (Eq. \ref{cm2s}), with $d_0$ obtained from CDFT
for the single surface. Hence, for the molecule attached to two metallic surfaces forming a molecular
junction, we approximate the overall corrections
for the molecular levels below $E_\mathrm{F}$ by
\begin{equation}
 \Sigma_\mathrm{o}(d)= - [I+\epsilon_{\mathrm{HOMO}}(d)] + U(d)\label{sigma_o}
\end{equation}
and similarly for the levels above $E_\mathrm{F}$ as
\begin{equation}
 \Sigma_\mathrm{u}(d)= -[A +\epsilon_{\mathrm{LUMO}}(d)] - U(d);\label{sigma_u}
\end{equation}
where $\epsilon_{\mathrm{HOMO/LUMO}}(d)$ is obtained from the position of the
peaks of the PDOS and $U(d)$ is the classical potential given by Eq. (\ref{cm2s}).
Here we assume a that the character of the molecular states is preserved when the junction is formed. 
\subsection{Electronic transport properties: DFT+NEGF}\label{negf_methods}
For the transport calculations, the system is divided into three regions: the central region, called scattering region
or device (D) region, which includes the molecule and a few layers of both the electrodes, and
the semi-infinite left (L) and right (R) electrodes, to which the device 
region is connected. The retarded Green's function of the device region, $\mathcal{G}_\mathrm{D}$, is then given by 
\begin{equation}
\mathcal{G}_{\mathrm{D}}(E)= \lim_{\eta\rightarrow 0}\left[(E+i\eta) -H_\mathrm{D}-\Sigma_\mathrm{L}-\Sigma_\mathrm{R}\right]^{-1} \label{eq:green-central},\\
\end{equation}
where $\Sigma_\mathrm{{L,R}}$ are the so-called
self-energies of the left-hand and right-hand side electrodes, $E$ is 
the energy and $H_\mathrm{D}$ is the KS-Hamiltonian
of the central region.
The electronic couplings between the electrodes and the device region are given by $ \Gamma_\mathrm{{L,R}}=i(\Sigma_\mathrm{{L,R}}-\Sigma_\mathrm{{L,R}}^{\dagger})$.
Following a self-consistent procedure,\cite{Rocha2006} 
the non-equilibrium  charge density extracted from Eq. (\ref{eq:green-central}) is used to calculate a new $H_{\mathrm{D}}[\rho]$.
Once the convergence is reached, the transmission coefficients are calculated as
\begin{equation}
 T=\mathrm{Tr}[\mathrm{\Gamma_\mathrm{L} \mathcal{G}_\mathrm{D}^\dagger\Gamma_\mathrm{R} \mathcal{G}_\mathrm{D}}]. \label{eq:trc}
\end{equation}
In the limit of zero-bias, we obtain the zero-bias conductance from the Fisher-Lee relation $G=G_0T(E_\mathrm{F})$
and the projected DOS (PDOS) for any orbital with index $\beta$ as
\begin{equation}
 \mathrm{PDOS}_\beta(E)=\frac{1}{2\pi}\mathrm{Im}[\mathcal{G}_{\mathrm{D}}(E)S_{\mathrm{D}}]_{\beta\beta}\label{eq:pdos}.
\end{equation}
In order to obtain reliable values for $T$ it is important to have an electronic structure theory capable of
describing the correct positions of the molecular energy levels with respect to $E_\mathrm{F}$,
since these ultimately dictate the transport properties of the device.
\section{RESULTS}
\subsection{Stability study of thiol-terminated molecules on a Au(111) flat surface and the junctions}\label{BDT-adsorption}

\begin{figure}[h]
\center
\includegraphics[width=0.46\textwidth]{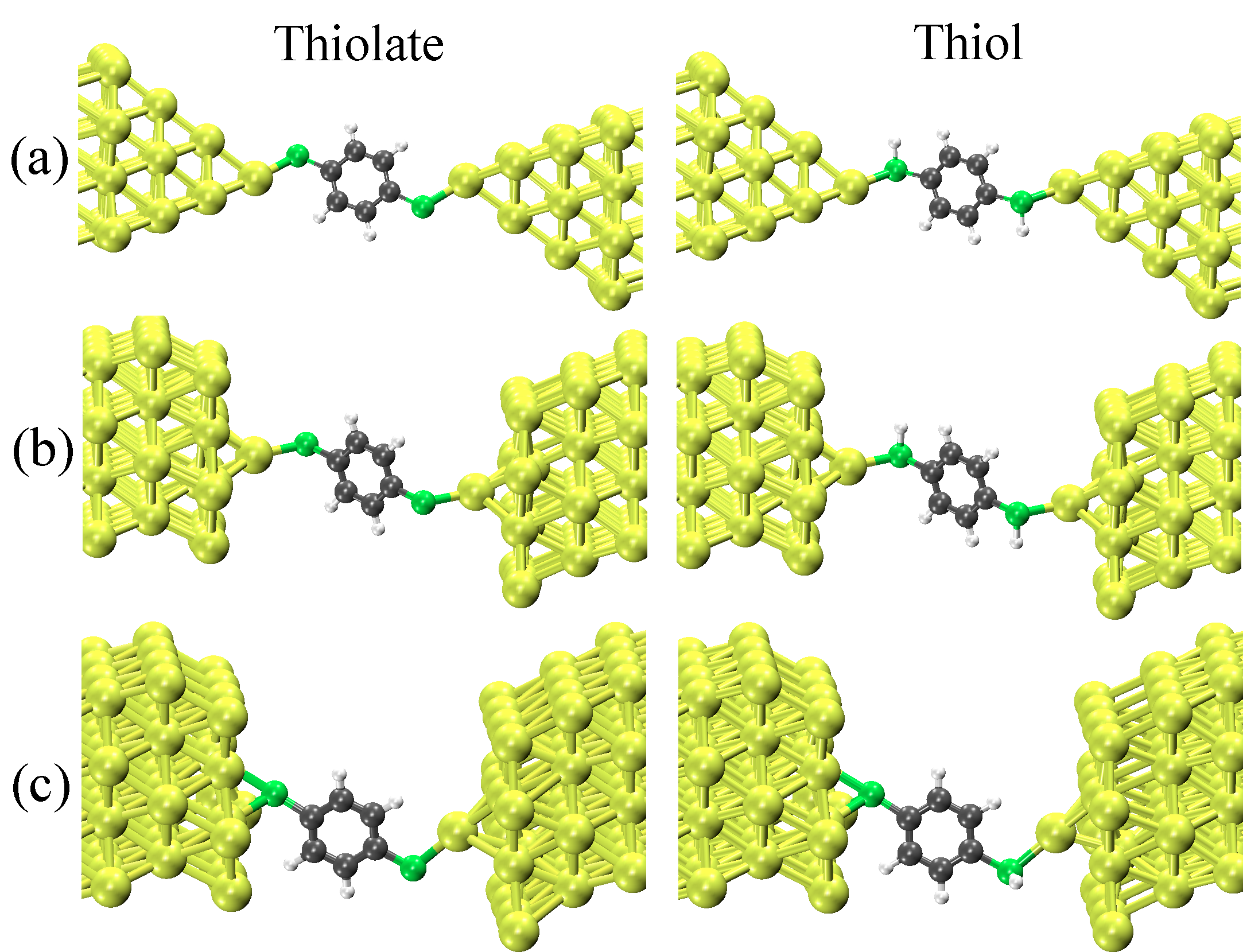}
\caption{Ball-stick representation of three molecule-electrode contact geometries. (a), (b) and (c) shows the tip-tip, 
adatom-adatom and surface-adatom configurations,
respectively. Left (right) panel shows the thiolate (thiol) junctions.}
\label{fig:junction-structures}
\end{figure}

In this section we present a systematic study, by means of total energy DFT calculations, of the stability of thiol-terminated
molecules on Au(111) flat surfaces, as well as when the molecule is attached to two Au electrodes forming a molecular junction.
For the systems presented in this section, the gold surface is modeled by considering a 3$\times$3 surface
unit cell five-layer thick.
This corresponds to a surface coverage of 1/3.\cite{Pontes2011a,Nara2004} The three bottom layers of
gold are kept fixed during the relaxation. For the junctions shown
in Fig. \ref{fig:junction-structures} we use a
slightly larger 4$\times$4 surface unit cell, in order to be able to model the tip-tip-like contact as well.

\begin{figure}[!t]
\center
\includegraphics[width=0.45\textwidth]{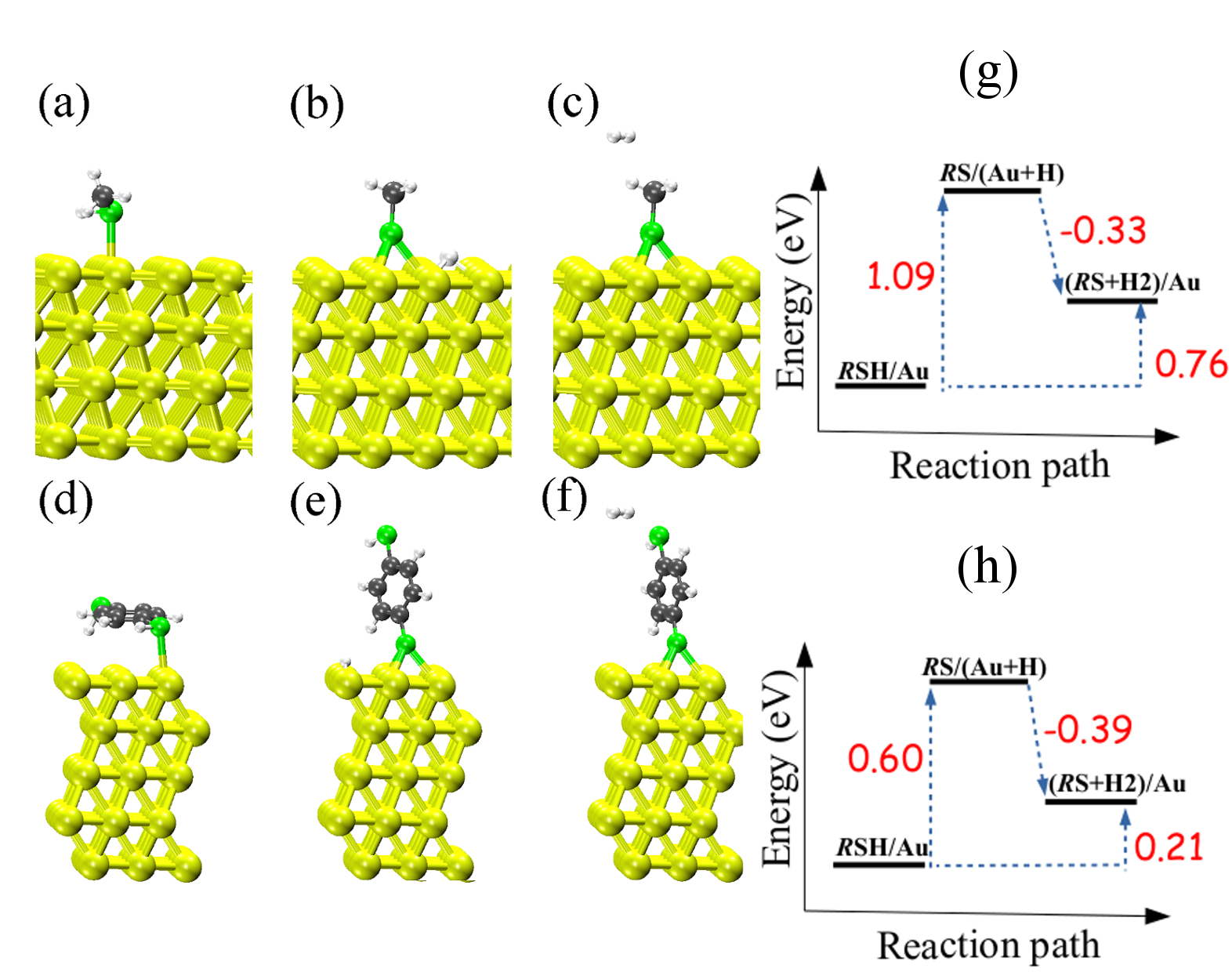}
\caption{Ball-stick representation of the adsorption process of methanethiol (a-c) and BDT2H (d-f)
on a flat Au(111) surface.
(a) and (d) the thiol molecules (\textit{R}SH/Au) are adsorbed on the surface;
(b) and (e) the hydrogen atom is dissociated to form thiolates (\textit{R}S/(Au+H)).
Finally, in (c) and (f) the hydrogen atoms attached to the Au surface desorbs
to form a $H_2$ molecule ((\textit{R}S+$H_2$)/Au). (g) and (h) schematically show the
total energy differences between each step of the reaction.}
\label{methiol@Au}
\end{figure}

We first discuss the adsorption process of benzene-1-4-dithiol ($\mathrm{C_6H_6S_2}$) on the Au(111) flat surface,
and compare it to adsorption properties of methanethiol ($\mathrm{CH_3SH}$). These molecules represent two distinct
classes, namely, aromatic and linear hydrocarbon compounds, respectively.
From this point on, we refer to benzene-1-4-dithiol as BDT2H in order to distinguish it
from the benzene-1-thiolate-4-thiol $\mathrm{C_6H_5S_2}$ (BDT1H), and from benzene-1-4-dithiolate $\mathrm{C_6H_4S_2}$ (BDT).
The calculations are performed as follows: (i) a system with the molecule terminated by a thiol group ($R\mathrm{SH/Au}$),
where $R=\mathrm{CH_3}$ for the methanethiol and $R=\mathrm{C_6H_5S}$ for the BDT2H, is placed close to the Au(111)
surface and the geometry is relaxed. (ii) Then a second system is built where the molecule is now terminated by a
thiolate group and a H atom is attached to the surface ($R\mathrm{S/Au+H}$), and again the geometry is relaxed.
Fig. \ref{methiol@Au}(a-c) shows the relaxed structures for the dissociative adsorption of the methanethiol
molecule, and the analogous structures are shown for the BDT2H in Fig. \ref{methiol@Au}(d-f).
For the $R\mathrm{SH/Au}$ systems, the molecule is tilted with respect to its vertical axis perpendicular
to the surface, whereas for the $R\mathrm{S/(Au+H)}$ systems the molecule is upright sitting on a hollow-site.
Our relaxed geometries are in good agreement with literature.\cite{Gronbeck2000,Nara2004}
We have also calculated the binding energies, as given by 
\begin{equation}
 E_\mathrm{b}=E_\mathrm{T}(R\mathrm{SH/Au})-E_\mathrm{T}(\mathrm{Au})-E_\mathrm{T}(R\mathrm{SH}), \label{biding_energy}
\end{equation}
for the methanethiol
and methanethiolate molecules on the Au(111) surface, and we find 0.63 eV and 1.42 eV, respectively.
For the BDT2H we find 0.12 eV whereas for the BDT1H, $E_\mathrm{b}$ is equal to 1.53 eV.
Finally, we consider a third structure for which the H atom attached to the surface is released from the surface
to form a $\mathrm{H_2}$ molecule $(R\mathrm{S+H_2)/Au}$. The formation energy
of the thiolate structure with a H atom attached to the surface is given by 
\begin{equation}
 E_{\mathrm{f}}=E_\mathrm{T}{(R\mathrm{SH/Au})}-E_\mathrm{T}{(R\mathrm{S/(Au+H)})}.
\end{equation}
Similarly, the formation energy for the dissociative adsorption followed
by the formation of a $\mathrm{H_2}$ molecule is calculated by
\begin{equation}
 E_\mathrm{f}= E_\mathrm{T}{(R\mathrm{SH/Au})}+\frac{1}{2}E_\mathrm{T}(\mathrm{H_2})-E_\mathrm{T}((R\mathrm{S+H_2)/Au}).
\end{equation}
Fig. \ref{methiol@Au}(g) and Fig. \ref{methiol@Au}(h) schematically show the total energy
differences between each step of the dissociative adsorption of the methanethiol and BDT2H molecules.
For the methanethiol molecule, if the dissociative reaction is accompanied by the chemisorption of a
H atom on the surface, as in Fig. \ref{methiol@Au}(b), the thiolate structure is energetically
unfavorable by 1.09 eV, a result consistent with previous calculations by Zhou \textit{et al.}~\cite{Zhou2006}
and temperature-programmed desorption (TPD) experiments.\cite{Lee2005,Nuzzo1987} When the H atoms adsorbed
on the surface are detached to form $\mathrm{H_2}$ molecules as in Fig. \ref{methiol@Au}(c),
the thiolate system becomes more stable by 0.33 eV compared to the thiolate system with the
H atom attached to the surface. Overall, the dissociative reaction followed by the formation
of a $\mathrm{H_2}$ molecule is unfavorable by 0.76 eV.
For the BDT2H molecule, the thiolate with a H atom attached to the surface is unfavorable by 0.60 eV
compared to the thiol structure, in good agreement with the value of 0.4 eV reported in recent studies
by Ning \textit{et al.}.\cite{Ning2010} When the dissociative reaction is accompanied by the formation
of a $\mathrm{H_2}$ from the H atom attached to the surface, this reaction is exothermic by 0.39 eV.
As a result, the dissociative absorption of BDT2H molecules on Au(111) surface followed by the
desorption of $\mathrm{H_2}$ is unfavorable by 0.21 eV. This partially contradicts the results
obtained by Nara \textit{et al.},\cite{Nara2004} who found the dissociative reaction
accompanied by the H atom on the surface to be indeed unfavorable by 0.22 eV. However,
for the case where the reaction is followed by the formation of $\mathrm{H_2}$,
the system is further stabilized by 0.42 eV so that the thiolate system is more stable by $\sim$0.20 eV.
Overall our results show that for both classes of molecules the dissociative
reaction is always unfavorable when considering either the formation of $R\mathrm{S/(Au+H)}$ or $(R\mathrm{S+H_2)/Au}$ structures.
\begin{figure}[!h]
\center
\includegraphics[width=0.45\textwidth]{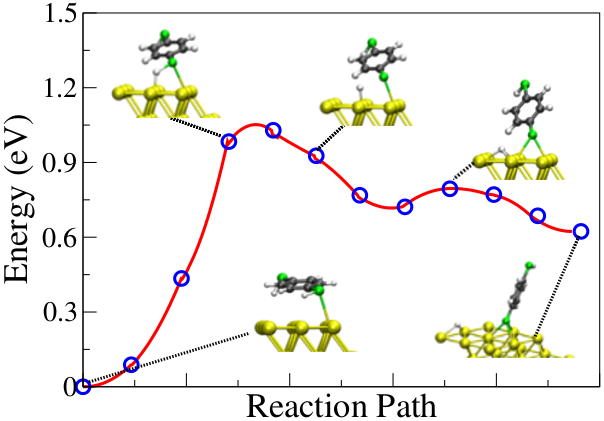}
\caption{Activation barrier for the dissociative adsorption of BDT2H on Au(111) surface as shown in Fig. \ref{methiol@Au}(d)-(e).}
\label{adsorption_bdt}
\end{figure}

In addition to the total energy differences between the dissociated and non-dissociated structures of BDT2H,
we evaluate the barrier height between those states [Fig. \ref{methiol@Au}(d) and Fig. \ref{methiol@Au}(e)], by means of
the Nudged Elastic Band (NEB) method,\cite{Henkelman2000, Henkelman2_2000, Henkelman3_2000} as shown in Fig. \ref{adsorption_bdt}. This allows us to estimate the transition probability between the states. Our results show that
the activation barrier is about 1 eV. The fact that the barrier is large provides evidence for possible existence of the thiol
structures on the surface, since a high temperature is required to overcome such a barrier. We note that defects
on the surface, such as adatom,
or the presence of a solvent, can change the energy barrier and eventually dissociation might take place at lower energies.

For the BDT2H molecule we also compare the stability of the thiol and thiolate structures when the molecule
is connected to two Au electrodes. We consider three types of junctions, as illustrated in Fig. \ref{fig:junction-structures}.
For the configuration shown in Fig. \ref{fig:junction-structures}(a),
ten gold atoms are added on each side of the junction forming a tip-like
symmetric contact with the molecule. For the configuration shown in Fig. \ref{fig:junction-structures}(b),
an adatom is added symmetrically on each side of the junction and for the one shown in Fig. \ref{fig:junction-structures}(c),
an adatom is added to one side of the junction and the molecule is connected to a flat surface on the other side.
These junctions constitute typical models for transport calculations found in the
literature.\cite{Toher2008,Strange2011a,Garcia-Lastra2009}
In this case, the formation energy difference between the thiol and the thiolate
structures with respect to the formation of $\mathrm{H_2}$ molecule is given by
\begin{equation}
E_\mathrm{f}= E_\mathrm{T}(\mathrm{BDT2H/Au})-E_\mathrm{T}(\mathrm{BDT/Au})-E_\mathrm{T}(\mathrm{H_2}),
\end{equation}
and the results are shown in Table. \ref{tab:formation_energy2}. 
Note that for the adatom-flat configuration the binding energy is evaluated considering $\frac{1}{2}\mathrm{H_2}$. 
For all the three junctions, the thiol
configurations are energetically more stable than their thiolate counterparts. 
\begin{table}[!h]
%\scalefont{0.85}
\begin{center}
\caption{Formation energy difference between the thiol and the thiolate
structures with respect to the formation of $\mathrm{H_2}$ molecule, in eV, for the three molecular junctions shown 
in Fig. \ref{fig:junction-structures}.}
{
\begin{tabular}{ccc }
\hline
\hline
System        &  VASP      & SIESTA             \\
\hline
surface-adatom&  -0.36      &  -0.42                \\
adatom-adatom &  -0.64      &  -0.40                \\
tip-tip       &  -0.77      &  -0.88                 \\
\hline
\hline
\end{tabular}\label{tab:formation_energy2}
}
\end{center}
\end{table}

\begin{figure}[!h]
\center
\includegraphics[width=0.43\textwidth]{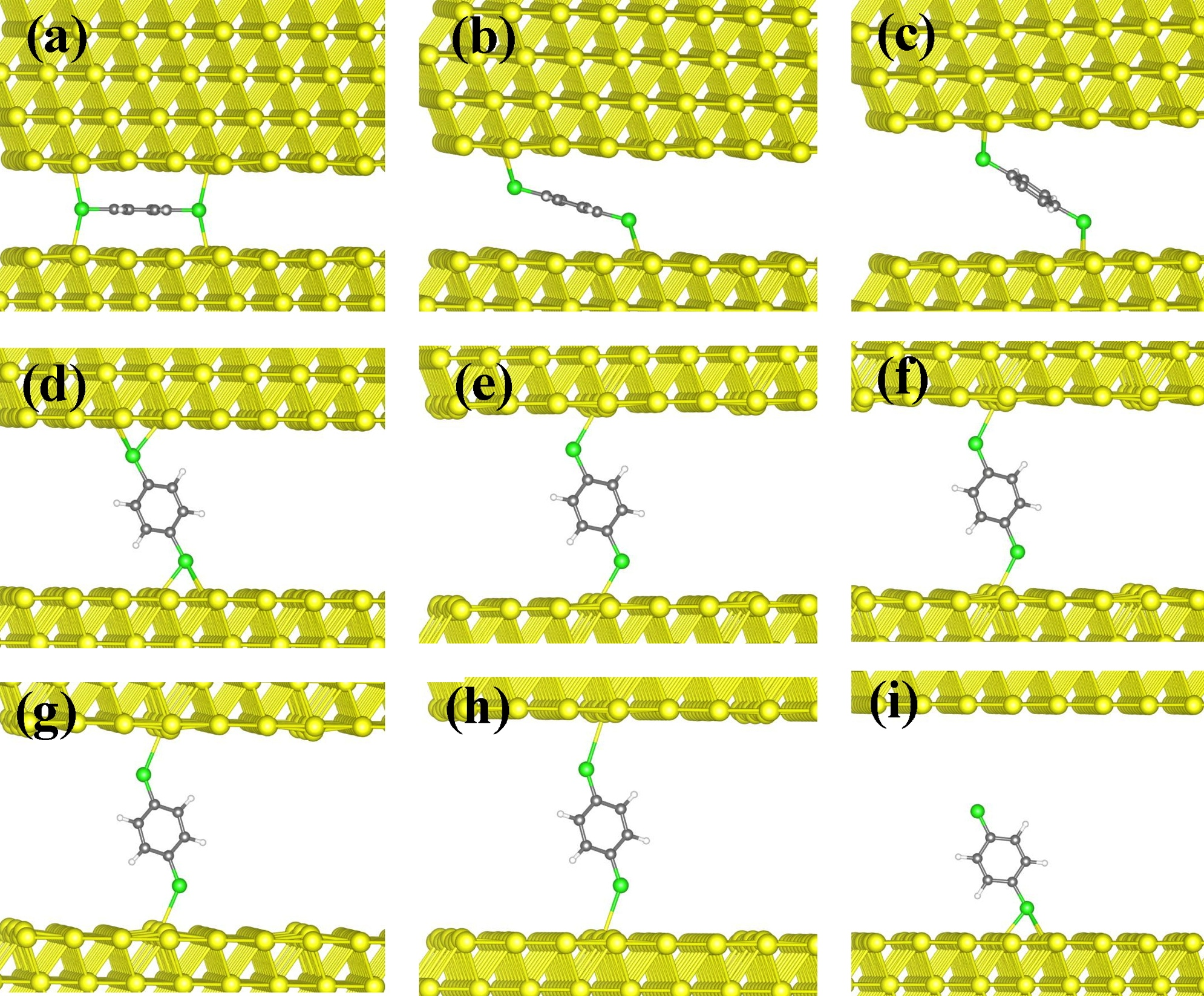}
\caption{(a)-(i) Ball-stick representation of the stretching process of BDT between two flat surfaces.}
\label{fig:bdtpics}
\end{figure}

\begin{figure}[!h]
\center
\includegraphics[width=0.42\textwidth]{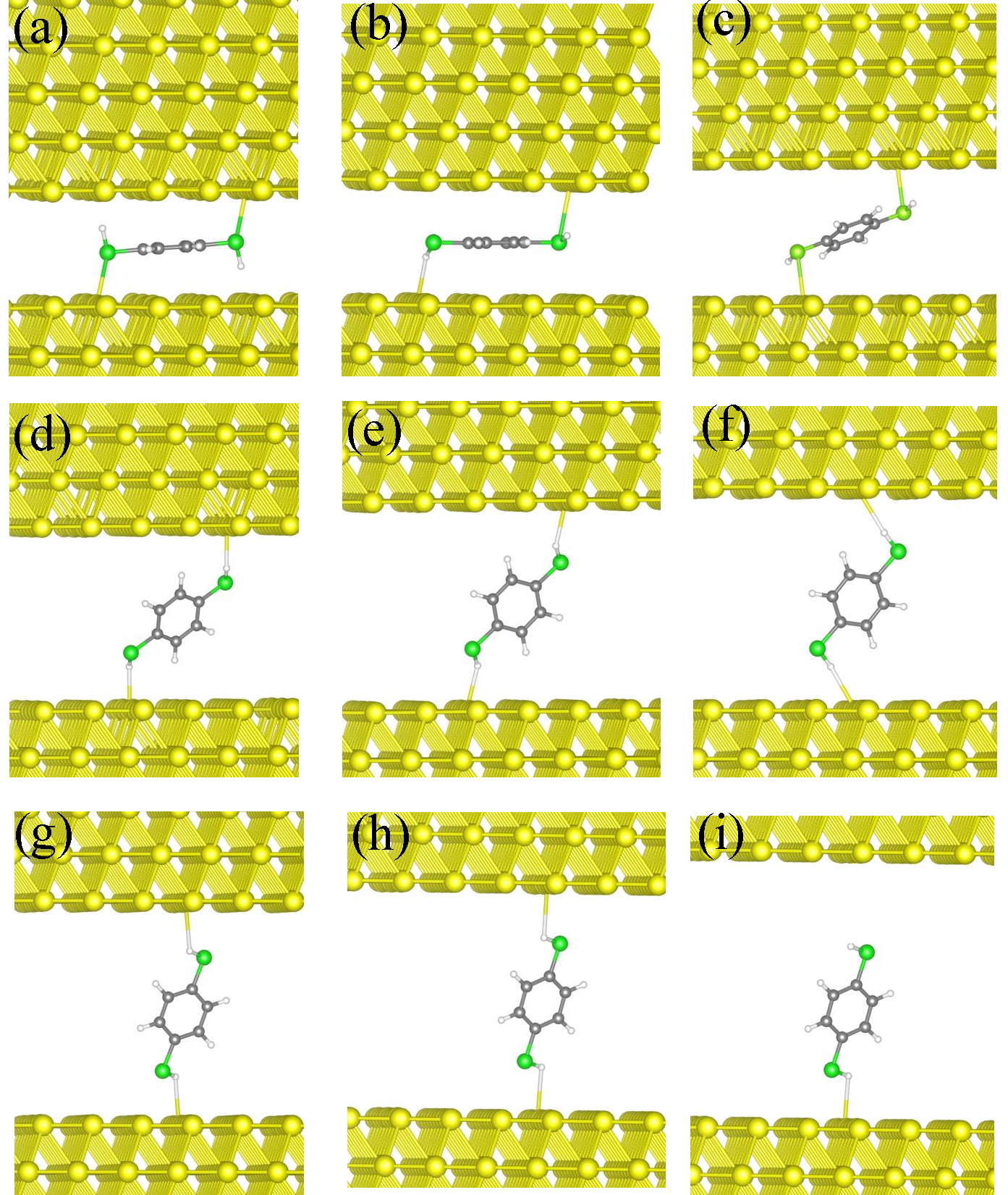}
\caption{(a)-(i) Ball-stick representation of the stretching process of BDT2H between two flat surfaces.}
\label{fig:bdt2hpics}
\end{figure}

One possibility that has been considered in order to determine whether there are thiols or
thiolates in the junction is a simultaneous measurement of $G$ and force in a STM and atomic
force microscopy (AFM) setup.\cite{Frei2012,Aradhya2012,Huang2006,Li2006} Since the binding
energy for thiol and thiolate can differ considerably, one might expect that the forces
involved when stretching the junction should be different. Therefore, we investigate
the energetics of Au(111)-BDT-Au(111) and Au(111)-BDT2H-Au(111) junctions as a function of $L$.
For the Au(111)-BDT-Au(111) junctions, similar calculations have been reported in the literature
in on attempt to simulate a MCBJ experiment within DFT.\cite{Pontes2011a,French2013, French2013a,
Strange2010, Romaner2006,Sergueev2010,Qi2009} Details on how the stretching is performed can be found
in Ref. [\onlinecite{Pontes2011a}]. Figs. \ref{fig:bdtpics}(a)-(i) and Figs. \ref{fig:bdt2hpics}(a)-(i)
show the relaxed structures for the Au(111)-BDT-Au(111) and Au(111)-BDT2H-Au(111) junctions undergoing stretching.

\begin{figure}[!h]
\subfigure{\includegraphics[width=0.23\textwidth]{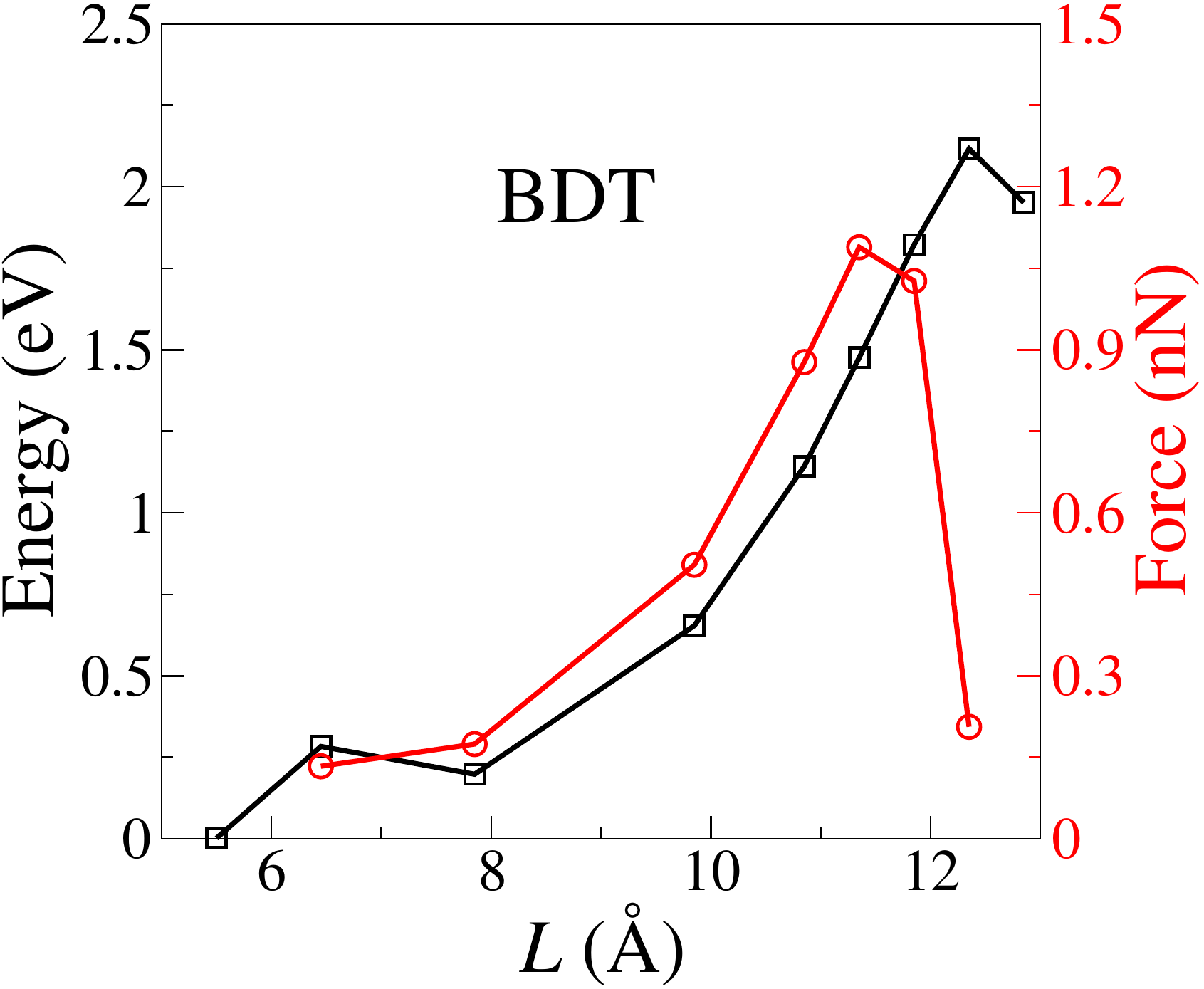}}
\hfill
\subfigure{\includegraphics[width=0.23\textwidth]{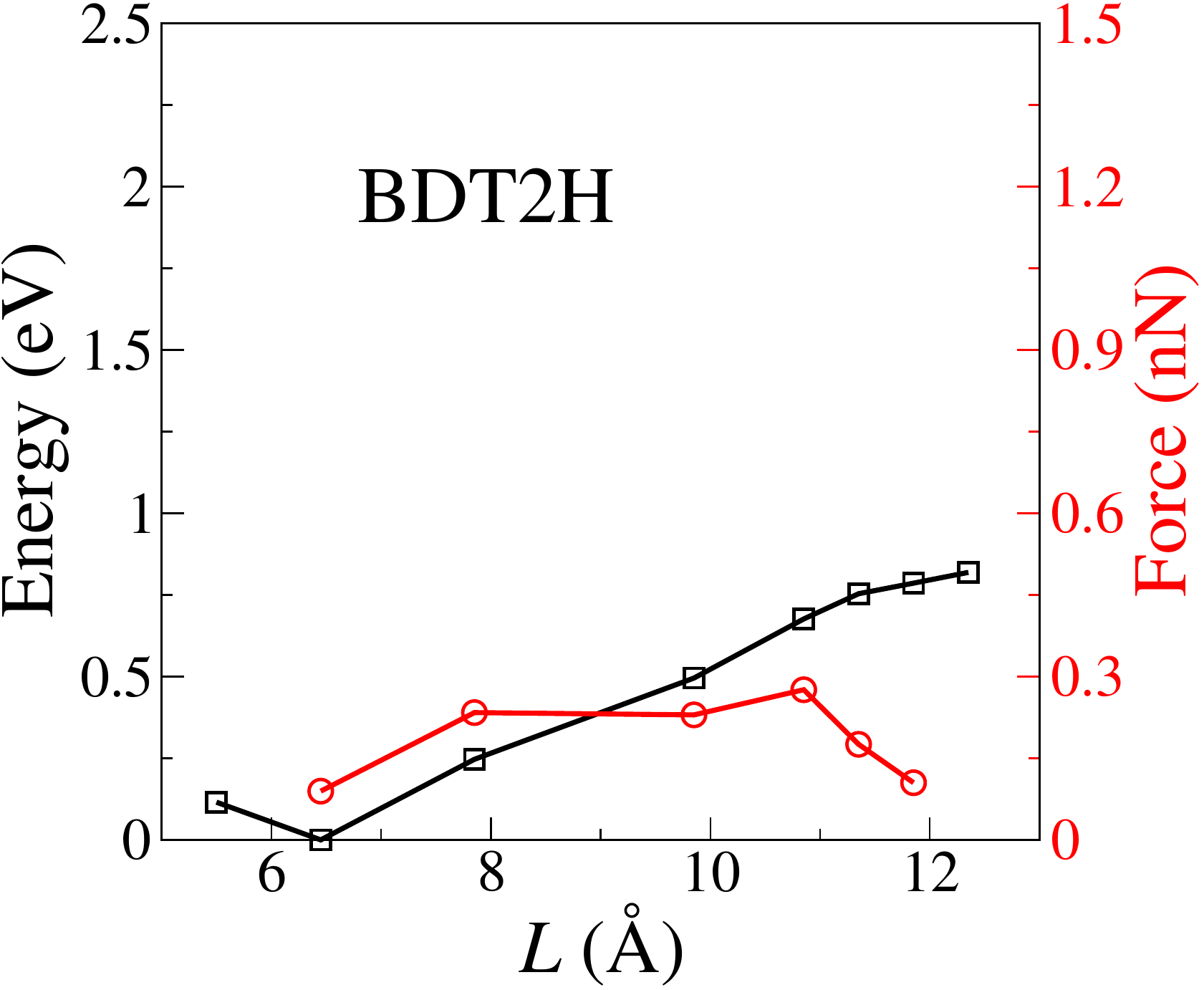}}
\caption{Total energy and pulling force as a function of $L$ for the Au-BDT-Au and Au-BDT2H-Au molecular junctions shown
in Fig. \ref{fig:bdtpics} and Fig. \ref{fig:bdt2hpics}, respectively.}
\label{fig:energy-forces}
\end{figure}

In Fig. \ref{fig:energy-forces} we show the energy and the forces as a function of $L$, for both Au(111)-BDT-Au(111)
and Au(111)-BDT2H-Au(111) junctions. Our results show that the breaking force for the S-Au
bond is about 1 nN, in good agreement with independent DFT results by
Romaner \textit{et al.}~\cite{Romaner2006} of 1.25 nN obtained using
the same contact geometry. The authors also considered the scenario when the BDT molecule is attached to an adatom contact geometry,
and they found that the breaking force can be as large as 1.9 nN.\cite{Romaner2006}
In fact, it is possible that during the elongation process the molecule is bonded to a single Au
atom rather than a flat surface.\cite{French2013}
For Au(111)-BDT2H-Au(111) our calculated breaking force is ~0.3 nN, as shown in Fig. \ref{fig:energy-forces}(b).
Thus the breaking forces for the BDT2H junctions are smaller than for those of BDT when the flat electrode geometry
is considered. We note that this is much smaller than the calculated value of 1.1-1.6 nN for the BDT2H molecule attached
to a tip-like contact geometry.\cite{Ning2010} 
Our small value of breaking forces of ~0.3 nN for the thiol junctions is consistent with the
rather small calculated $E_\mathrm{b}$ of 0.12 eV, and indicates weak coupling between the molecule and the flat electrodes.
A similar study for a octanedithiol-Au junction has also been reported,\cite{Qi2009} and for
an asymmetric junction they found the breaking force of the Au-thiol bond to be 0.4-0.8 nN. 
Other experiments using the same molecule~\cite{Huang2006,Li2006} reported a breaking
force of ~1.5 nN, which is very similar to the breaking force of a Au-Au bond,
therefore, leading to the conclusion that the junction might break at the Au-Au
bond and also indicating the presence of Au-thiolate instead of Au-thiol junctions.

In summary, we find that the dissociative reaction of methanethiol and BDT2H on Au(111)
is energetically unfavorable. Especially for the BDT2H, the activation barrier of $\sim$1
eV strongly suggests the presence of thiol structures when the molecules attach to the metallic
surface. Moreover, for all the contact geometries of molecular junctions presented in
Figs. \ref{fig:junction-structures}-\ref{fig:bdt2hpics}, the thiol systems are also
energetically more stable. These results indicate that the non-dissociated structures
are likely to exist in experiments, and therefore should be considered when modeling
transport properties of such systems.

\subsection{Energy level alignment}\label{level-alignment}

One of the possible reasons for the discrepancies between theory and experiments regarding the conductance
of molecular junctions is the difficulty, from a theoretical point of view, to obtain the correct energy
level alignment of such systems. Table. \ref{tab:eigenvalues_IP_EA} shows the LDA eigenvalues for the frontier
molecular states of BDT and BDT2H in the gas phase. $E_\mathrm{LDA}^\mathrm{gap}$ is largely underestimated
when compared to $E_\mathrm{QP}^\mathrm{gap}=I-A$ calculated by the so-called delta self-consistent
field ($\Delta$SCF) method. For the BDT molecule,
our results show that the HOMO is higher in energy by 2.73 eV with respect to $-I$ whereas the LUMO is lower
in energy by 2.66 eV compared to $-A$. For the BDT2H, the HOMO is higher in energy by 2.49 eV with respect
to $-I$, and the LUMO is lower in energy by 2.51 eV when compared to $-A$.
The results clearly indicate that the KS eigenvalues offer a poor
description of the molecule quasi-particle levels even in the gas phase within GGA/LDA.

\begin{figure}[!h]
\includegraphics[width=0.5\textwidth]{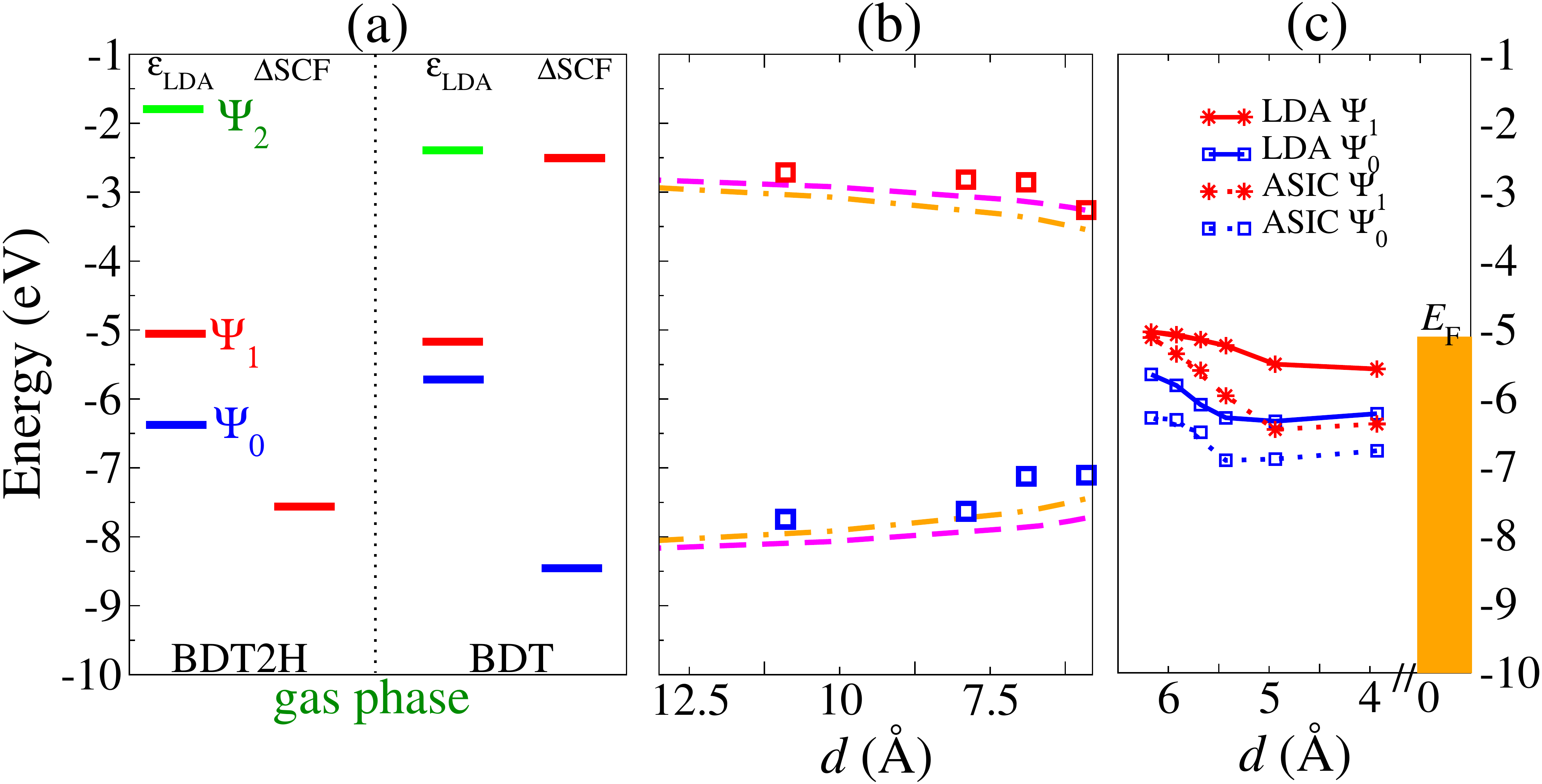}
\caption{Energy level alignment of the frontier molecular orbitals for the BDT molecule
from the gas phase to the formation of the Au-BDT-Au junction. (a) 
LDA eigenvalues ($\epsilon_\mathrm{LDA}$) and $\Delta$SCF calculations in the gas phase BDT.
For comparison, we also show results for the gas phase of BDT2H. All the values are given with respect to the vacuum level.
(b) CDFT calculations for
the charge-transfer energies between BDT molecule adsorbed and a single flat surface: $E_\mathrm{CT}^{+}(d)$ (blue squares) and $E_\mathrm{CT}^{-}(d)$ (red squares).
The classical image charge contribution for two surfaces (dashed-line) and for a single surface
(dashed-dotted line) are plotted for comparison where $E_\mathrm{F}=-5.1$ eV and $d_0=1$~\AA.
(c) LDA and ASIC energy
levels for the HOMO ($\Psi_0$) and LUMO ($\Psi_1$) obtained from the PDOS peaks for
the molecule at the junction as a function of $d=L/2$.}
\label{fig:energy-alignment}
\end{figure}

\begin{figure}[!h]
\center
\includegraphics[width=0.45\textwidth]{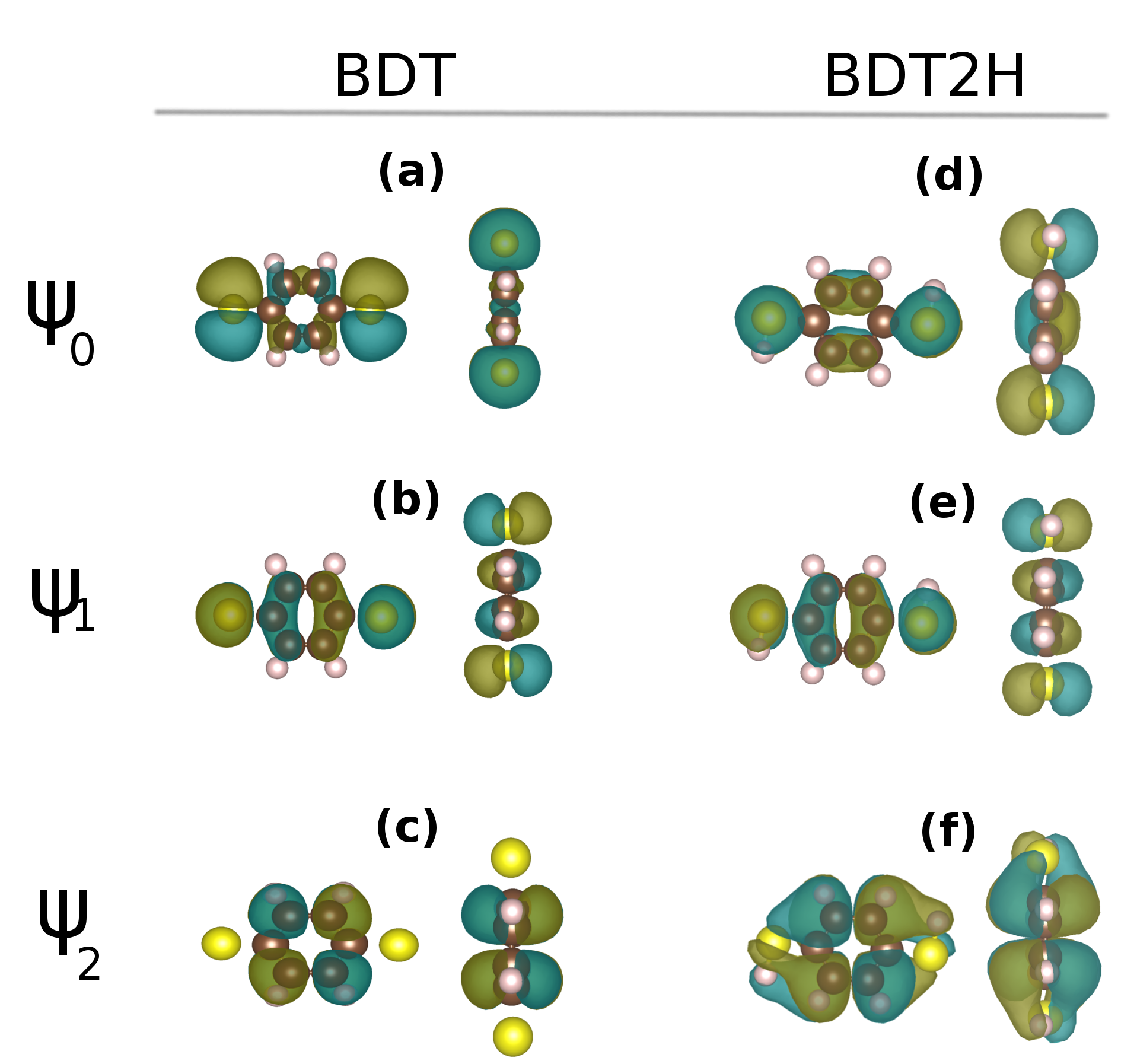}
\caption{Plots of wavefunctions: (a), (b) and (c) show $\Psi_0$ (20th state),
$\Psi_1$ (21st state) and $\Psi_2$ (22sd state), respectively, for the gas phase
BDT molecule; (d), (e) and (f) show the same for the the BDT2H molecule. Isosurfaces are taken at 
a density of 0.06 $e/\mathrm{\AA^3}$.}
\label{fig:homo-lumo-bubles}
%\end{minipage}
\end{figure}

Fig. \ref{fig:energy-alignment}(a) shows schematically the energies of these states for the gas phase molecules.
In the case of BDT2H molecule, the wavefunctions $\Psi_0$ (blue), $\Psi_1$ (red) and $\Psi_2$ (green) correspond to the HOMO-1,
HOMO and LUMO of the isolated molecule, respectively. For the BDT the removal of 2 H atoms when compared to BDT2H leads to a
reduction of the number of electrons by 2 as well, so that to a first approximation the BDT2H HOMO becomes the LUMO for the
BDT molecule [see Fig. \ref{fig:homo-lumo-bubles}(b) and \ref{fig:homo-lumo-bubles}(e)]. Therefore, for BDT,
$\Psi_0$ corresponds to the HOMO, $\Psi_1$ to the LUMO, and $\Psi_2$ to the LUMO+1.
Fig. \ref{fig:homo-lumo-bubles} shows the real space representation
of $\Psi_0$, $\Psi_1$ and $\Psi_2$ for BDT (left) and BDT2H (right) molecules in the gas phase. 
{
\begin{table}[!h]
\scalefont{1.0}
\begin{center}
\caption{Calculated LDA eigenvalues ($\epsilon$), $E_\mathrm{LDA}^\mathrm{gap}$,
$I$, $A$ and $E_\mathrm{QP}^\mathrm{gap}$ (calculated with $\Delta$SCF) for the gas phase
BDT and BDT2H molecules.}
{
\renewcommand{\arraystretch}{1.3}
\setlength{\tabcolsep}{2pt}
\begin{tabular}{lccccccc }
\hline
\hline
           & \multicolumn{3}{c}{LDA} &   & \multicolumn{3}{c}{$\Delta$SCF}   \\
           \cline{2-4}\cline{6-8}
  System    & $\epsilon\mathrm{_{HOMO}}$&$\epsilon\mathrm{_{LUMO}}$& $E_\mathrm{LDA}^\mathrm{gap}$ && -I & -A &$E_\mathrm{QP}^\mathrm{gap}$ \\
\hline
BDT        & -5.74  &  -5.19   &  0.55 && -8.47 & -2.53 & 5.94  \\
BDT2H      & -5.09  &  -1.82   &  3.27 && -7.58 & 0.69  & 8.27  \\ 
\hline 
\hline
\end{tabular}\label{tab:eigenvalues_IP_EA}
}
\end{center}
\end{table}
}

Fig. \ref{fig:energy-alignment}(b) shows the CDFT results for $E_{\mathrm{HOMO}}/E_{\mathrm{LUMO}}$ as
a function of $d$ for the BDT/Au(111) system (see Sec. \ref{cdft_methods}). In the CDFT calculations
the metal is modeled by a 9$\times$9 Au(111) surface with five atomic layers and the molecule placed upright
at a distance, $d$, from the center of the molecule to the Au surface, and we use a 20~\AA~vacuum
region in the direction perpendicular to the surface plane.
We note that although CDFT is in principle applicable at all $d$, when $d$ becomes less than about $5.9$~\AA,
for which the Au-S bond distance, $d_\mathrm{Au-S}$, is less than 2.5~\AA, the amount of charge on each fragment
is ill defined due to the hybridization between the molecular orbitals and the electrode continuum.
Therefore, at those small distances, the CDFT charge-transfer energies are not well defined. At $d=5.9$~\AA,
the CDFT calculations give an overall reduction of $E_\mathrm{QP}^\mathrm{gap}$
of 2.09 eV with respect to
the value obtained for isolated BDT.
\begin{figure}[!h]
\center
\includegraphics[width=0.45\textwidth]{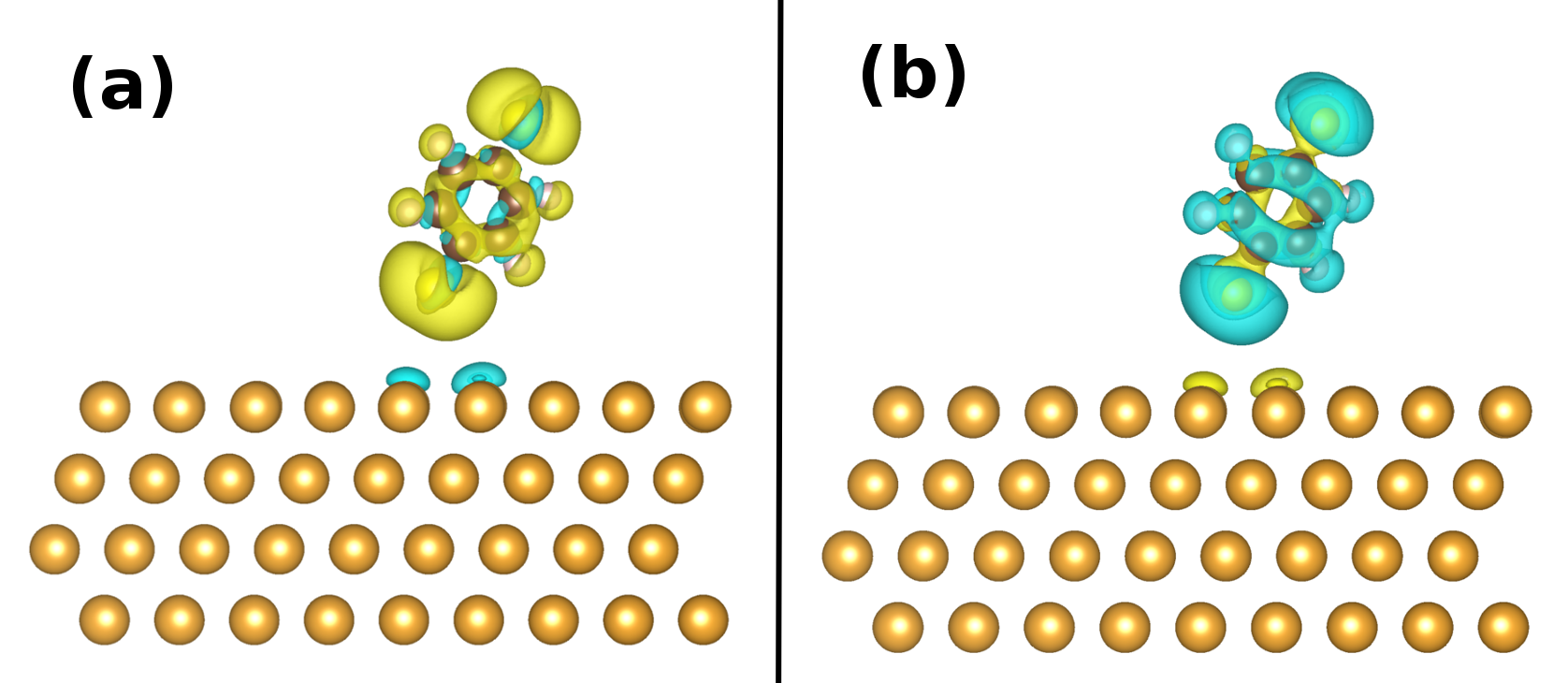}
\caption{Charge density differences for (a) $E_\mathrm{CT}^{+}(d)$ and (b) $E_\mathrm{CT}^{-}(d)$ for $d=6.9$~\AA. Isosurfaces are 
taken at $10^{-4}~\mathrm{e/\AA^3}$.}
\label{fig:cdft-bubles}
%\end{minipage}
\end{figure}
Fig. \ref{fig:energy-alignment}(b) also shows the results of the classical model for the image charge
calculated for one [Eq. (\ref{cm1s}), dashed line] and for two [Eq. (\ref{cm2s}), dash-dotted line] surfaces.
The CDFT $d_0$ ranges from 0.79~\AA~to 1.13~\AA, depending on the distance,
and we therefore take $d_0=1$~\AA~as average value. Coincidentally, this is the
same value used in literature,\cite{Garcia-Suarez2011,Quek2011,Perrin2013}
where it was however not formally justified, but rather used as a free parameter.
The corrections to $I$ and $A$ from the classical model when considering two surfaces
are larger than the corrections for a single surface, since $U(d)>V(d)$ for all $d$.
We evaluate the charge density differences between the constrained and non-constrained
calculations for $E_\mathrm{CT}^{+}(d)$ and $E_\mathrm{CT}^{-}(d)$ (Fig. \ref{fig:cdft-bubles}).
It can be seen that the hole (electron) left on the molecule has the same character as the
corresponding $\Psi_0$ ($\Psi_1$) wavefunction [compare to Figs. \ref{fig:homo-lumo-bubles}(a)-(b)].  

Fig. \ref{fig:energy-alignment}(c) shows the energies of the eigenvalues of the $\Psi_0$ and $\Psi_1$ states for the BDT
molecule as a function of $d=L/2$, calculated with LDA (solid lines) and ASIC (dashed lines), for the stretching
configurations shown in Fig. \ref{fig:bdtpics}(c-h). The energies of these levels are set to be at 
the peaks of the corresponding PDOS. In the limit of weak coupling between the BDT molecule and the electrodes,
which is the case for $L=12.35$~\AA, at which $d_\mathrm{Au-S}$ is the largest before
rupture of the junction, LDA gives the LUMO of the isolated BDT molecule ($\Psi_1$) slightly
above $E_\mathrm{F}$. However, as shown in Fig. \ref{fig:energy-alignment}(a) and Table. \ref{tab:eigenvalues_IP_EA},
the corrected energy of $\Psi_1$ (which is given by $-A$) is 2.66 eV above the LDA eigenvalue.
Similarly, the LDA energy of $\Psi_0$ is too high by 2.73 eV when compared to $-I$.
The same analysis can be done for $L=11.86$~\AA~and $L=11.36$~\AA,
for which $\Psi_1$ is still above $E_\mathrm{F}$.
In other words, for $L\ge11.36$\AA, 
the molecule is weakly bonded to the electrodes, therefore, charge transfer from the electrodes to the molecule due to the hybridization
of the molecular and electrodes states is small.
These results show that for the Au-BDT-Au junctions in the weak coupling regime the
LDA BDT HOMO (corresponding to $\Psi_0$) is in fact too high in energy whereas
the LDA BDT LUMO (corresponding to $\Psi_1$) is too low.

In order to correct the energy levels,
we apply the SCO method (see Sec. \ref{sco_methods}). Table. \ref{bdt2h_energy_corrections} shows,
for the Au-BDT2H-Au and Au-BDT-Au junctions, $U$ [Eq. (\ref{cm2s})], $\Sigma_\mathrm{o}$ [Eq. (\ref{sigma_o})]
and $\Sigma_\mathrm{u}$ [Eq. (\ref{sigma_u})] as functions of $L$. 
As pointed out by Garcia-Suarez \textit{et al.},\cite{Garcia-Suarez2011} the shift of the energy
level is unambiguous when there is no resonance at $E_\mathrm{F}$,\cite{Quek2007,Quek2011} so that 
the occupied levels are shifted downwards and the empty levels are shifted upwards in energy.
This is the case for the BDT2H molecule, where the isolated molecule has 42 electrons,
therefore the 21st molecular level is the HOMO of the isolated molecule ($\Psi_1$ in this case).
Since it is already filled with two electrons, it lies below $E_\mathrm{F}$ when the molecule is
in the junction, and the LUMO ($\Psi_2$) is always empty and well above $E_\mathrm{F}$.
\begin{table}[!h]
\scalefont{0.9}
\begin{center}
\caption{Contribution due to the classical image charge
for two surfaces model ($U$) and
the final corrections $\Sigma_\mathrm{o}/\Sigma_\mathrm{u}$ as a function of $L$ for BDT2H and BDT molecules at the junction. The first column 
correspond to the labels of a subset of the structures shown in Fig. \ref{fig:bdtpics} and Fig. \ref{fig:bdt2hpics}. The Au-S bond distance,
$d_\mathrm{Au-S}$, is also shown for completeness.}
{
\renewcommand{\arraystretch}{1.0}
\setlength{\tabcolsep}{1.5pt}
\begin{tabular}{ccccccccccccc }
%\begin{tabular}{c c c c }
\hline
\hline

                            &&&&&                                & \multicolumn{3}{c}{BDT2H} & \multicolumn{3}{c}{BDT}\\
                            \cline{7-8}\cline{10-12}
&$L$ (\AA)  & $d_\mathrm{Au-S}$ (\AA) && U (eV)&  &$\Sigma_\mathrm{o}$(eV)  & $\Sigma_\mathrm{u}$ (eV)  & & $\Sigma_\mathrm{o}$(eV) && $\Sigma_\mathrm{u}$ (eV)  \\
\cline{1-3}\cline{4-5}\cline{7-8}\cline{10-12}
(c)&7.86         & 2.11              && 1.70  &    &  0.18       & 1.63              & & -  &&  -                       \\
(d)&9.89         &  2.08             && 1.26  &    &  -0.50       & 1.72              & &-   &&  -                   \\
(e)& 10.87         & 2.47            & & 1.12  &    &  -0.80       & 1.78              & & - & &  -                    \\
(f)& 11.36         & 2.67            && 1.06  &    &  -1.04       & 1.77              & &  -1.13       && 1.45                    \\
(g)& 11.84        & 2.90            && 1.01  &    &  -0.62       & 2.22              & &  -1.63       && 1.53                      \\
(h)& 12.35         & 3.18          &  & 0.96  &    &  -0.83       & 2.17              &  &  -1.84      & & 1.55                    \\
%$\infty$          & 0     &    &  -2.18     & 1.77      &  & -     &   -                     \\ 

\hline
\hline
\end{tabular}\label{bdt2h_energy_corrections}
}
\end{center}
\end{table}

For closer distances, due to the stronger coupling between molecule and electrodes,
hybridization occurs leading to a fractional charge transfer from the electrodes to the molecule.
For small $d$ also for the BDT molecule the $\Psi_1$ state becomes partially occupied,
and positioned slightly below $E_\mathrm{F}$. This means that, for the structures
considered in Fig. \ref{fig:bdtpics}, the correction defined by Eq. \ref{sigma_u}
can not be applied for $L\le10.87$, since this is the distance where the level moves slightly below $E_\mathrm{F}$.
We note that when the level is pinned at $E_\mathrm{F}$, many-body effects become
important, and the GW method might be the most appropriate approximation.\cite{Strange2011}
Once the coupling is strong enough, and $\Psi_1$ is almost fully filled, it becomes effectively
the HOMO of the BDT. In this case we expect its energy to be too high within LDA, and therefore application
of ASIC is expected to improve its position with respect to $E_\mathrm{F}$. 
In fact, ASIC corrects $\Psi_1$ by $\sim$1 eV as $d$ decreases, as shown in Fig. \ref{fig:energy-alignment}(c).
%For instance for $\Psi_1$,
%as shown in Fig. \ref{fig:energy-alignment}(a), ASIC shifts the level by -0.8 eV, -0.95 eV and -0.73 eV
%for $L=7.86$~\AA, $L=9.89$~\AA~ and $L=10.87$~\AA, respectively. 

For the weak coupling limit the calculated corrections show that $\Psi_1$ (the LUMO of the isolated BDT molecule)
is empty and its LDA eigenvalue is too low in energy. In contrast, for the strong coupling limit the energy
of $\Psi_1$ moves below $E_\mathrm{F}$, so that the state becomes occupied, and its LDA eigenvalue
is now too high in energy. In this regime we apply the ASIC method to give a better description
of the energy level alignment.

\subsection{Electronic Transport Properties: Thiol versus Thiolate Junctions}\label{Transport}
For the electronic transport properties, we start by presenting results for the
molecular junctions at fixed distance and different molecule-surface bonding. Subsequently we discuss
the conductivity of the thiol and thiolate systems attached to flat Au electrodes under stretching. 
\begin{figure}[!h]
\center
\subfigure{\includegraphics[width=0.45\textwidth]{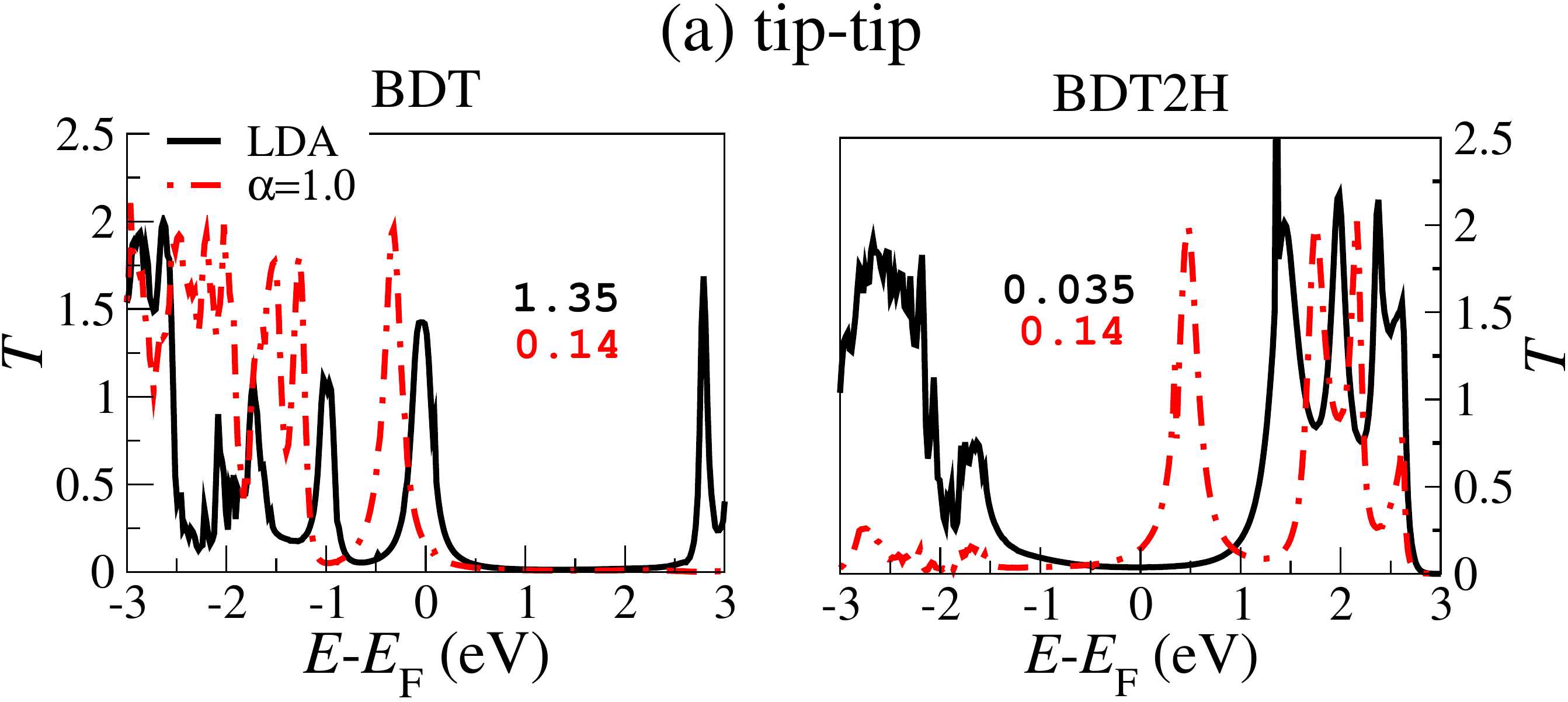}}
\subfigure{\includegraphics[width=0.45\textwidth]{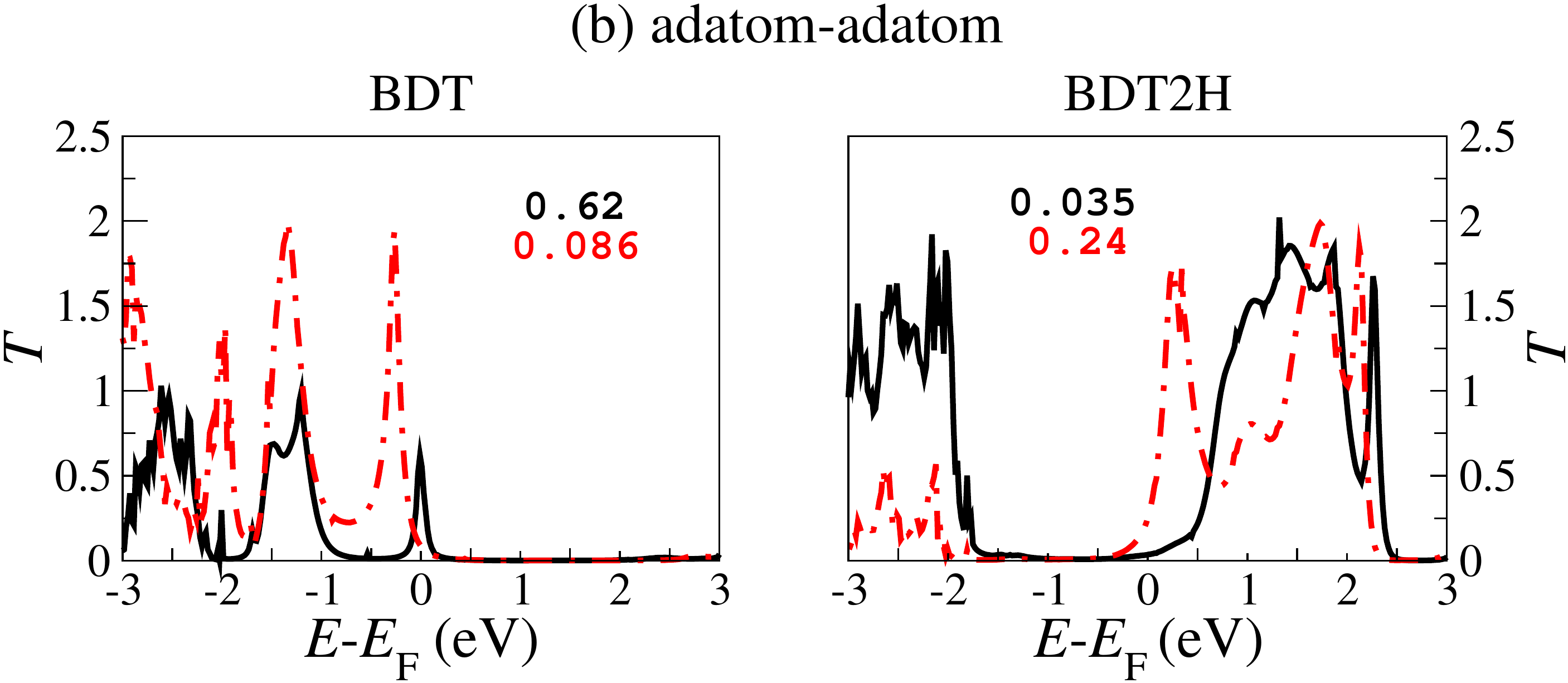}}
\subfigure{\includegraphics[width=0.45\textwidth]{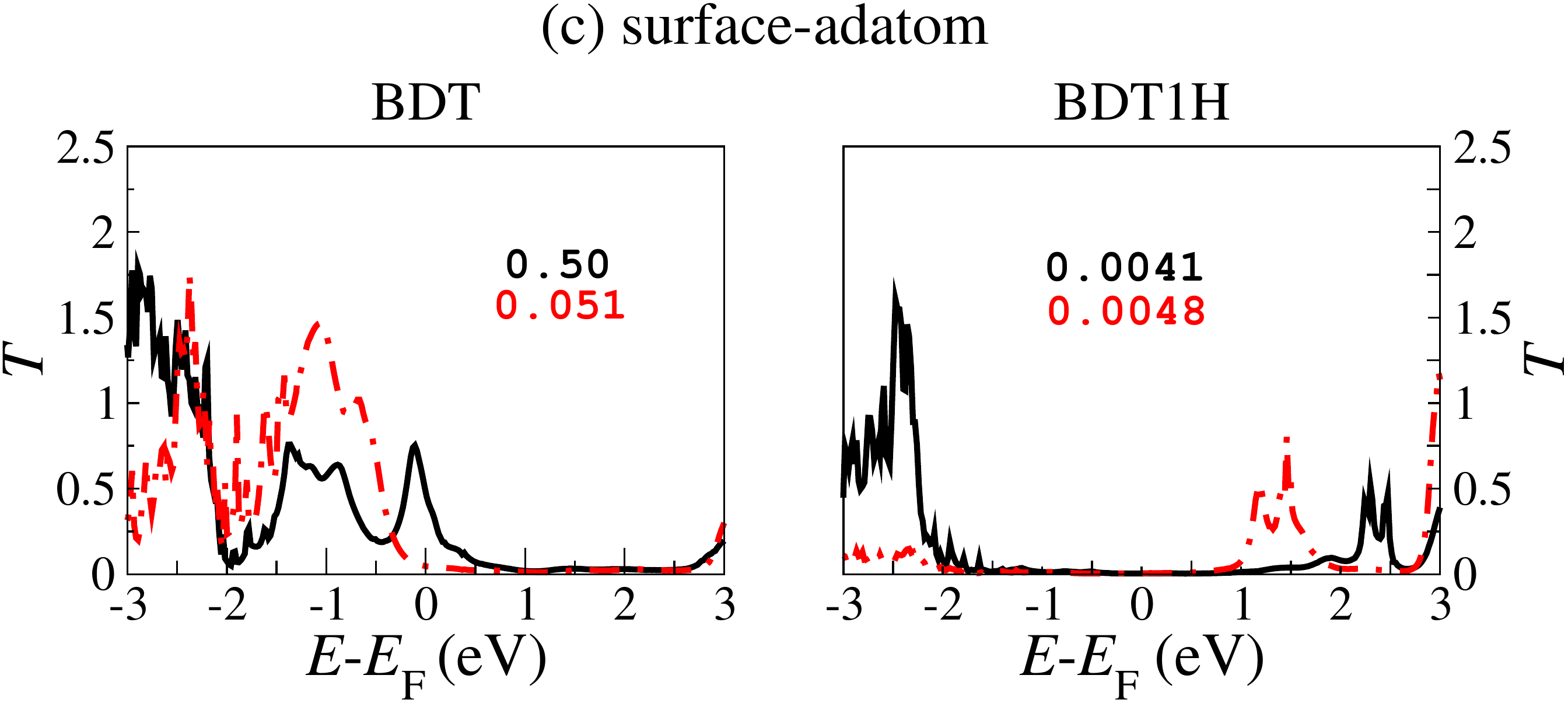}}
\caption{(a) Transmission coefficients as a function of energy for thiolate (left column) and 
thiol (right column) for the structures shown in Fig. \ref{fig:junction-structures}. 
For each case we report the transmission at $E_\mathrm{F}$ for both LDA (black full-line) and ASIC (red dashed-line)
results.}
\label{fig:trc-asic}
\end{figure}

Fig. \ref{fig:trc-asic} shows $T(E)$ for the thiolate (left column) and thiol (right column),
for \textit{tip-tip}, \textit{adatom-adatom} and \textit{surface-adatom} structures
(see Fig. \ref{fig:junction-structures} for the structure geometries).
Within the LDA functional, the transmission curves of all the thiolate
junctions present a peak pinned at $E_\mathrm{F}$. These results have
been found in several works reported in the literature for Au-BDT-Au
(thiolate) junctions.\cite{Toher2008,Pontes2011a,Strange2011,Kim2011,French2013,French2013a}
The resonant states at $E_\mathrm{F}$ yield high values of $G$ of 1.35$G_0$, 0.45$G_0$ and 0.22$G_0$
for \textit{tip-tip}, \textit{surface-adatom} and \textit{adatom-adatom}, respectively.
The observed peaks at $E_\mathrm{F}$ correspond to the hybridized $\Psi_1$ state of the BDT molecule.
Note that the exact position of the peaks and so the exact $G$ values depend on the atomistic
details of the junctions, as well as on the functionals used within DFT. 
We point out that such high values of $G$ have never been observed experimentally,
indicating that LDA does not give the correct energy level alignment between the molecule and the electrodes,
as already discussed in Sec. \ref{level-alignment}. 
In contrast, for the thiol junctions, no resonant states are found around $E_\mathrm{F}$.
The zero-bias conductance is in the range of 0.035-0.004$G_0$, which is in good agreement
with experimental values of 0.011$G_0$.\cite{Tao2004,Bruot2012,Kiguchi2010,Tsutsui2009a}

When ASIC is used, for the BDT structures the molecular energy level remains pinned
at $E_\mathrm{F}$, and it is just slightly shifted to lower energies. This slight
shift is however enough to decrease $G$ by one order of magnitude. For the hydrogenated
junctions (BDT1H and BDT2H) there are no molecular states at $E_\mathrm{F}$ for LDA,
and in this case ASIC shifts downwards the energy levels of the occupied states. We also note that the empty 
states are shifted down in energy, which is an artifact of the ASIC method, as discussed in Sec. \ref{asic_methods}.
This shows that, while the ASIC method improves the position of the levels below $E_\mathrm{F}$,
it can lead to down-shifts for the empty states, resulting in a spurious enhanced $G$ due to the LUMO.
Further corrections are therefore needed in order to give a quantitatively correct value of $G$ in such systems. 

\begin{figure}[!h]
\center
\includegraphics[width=0.46\textwidth]{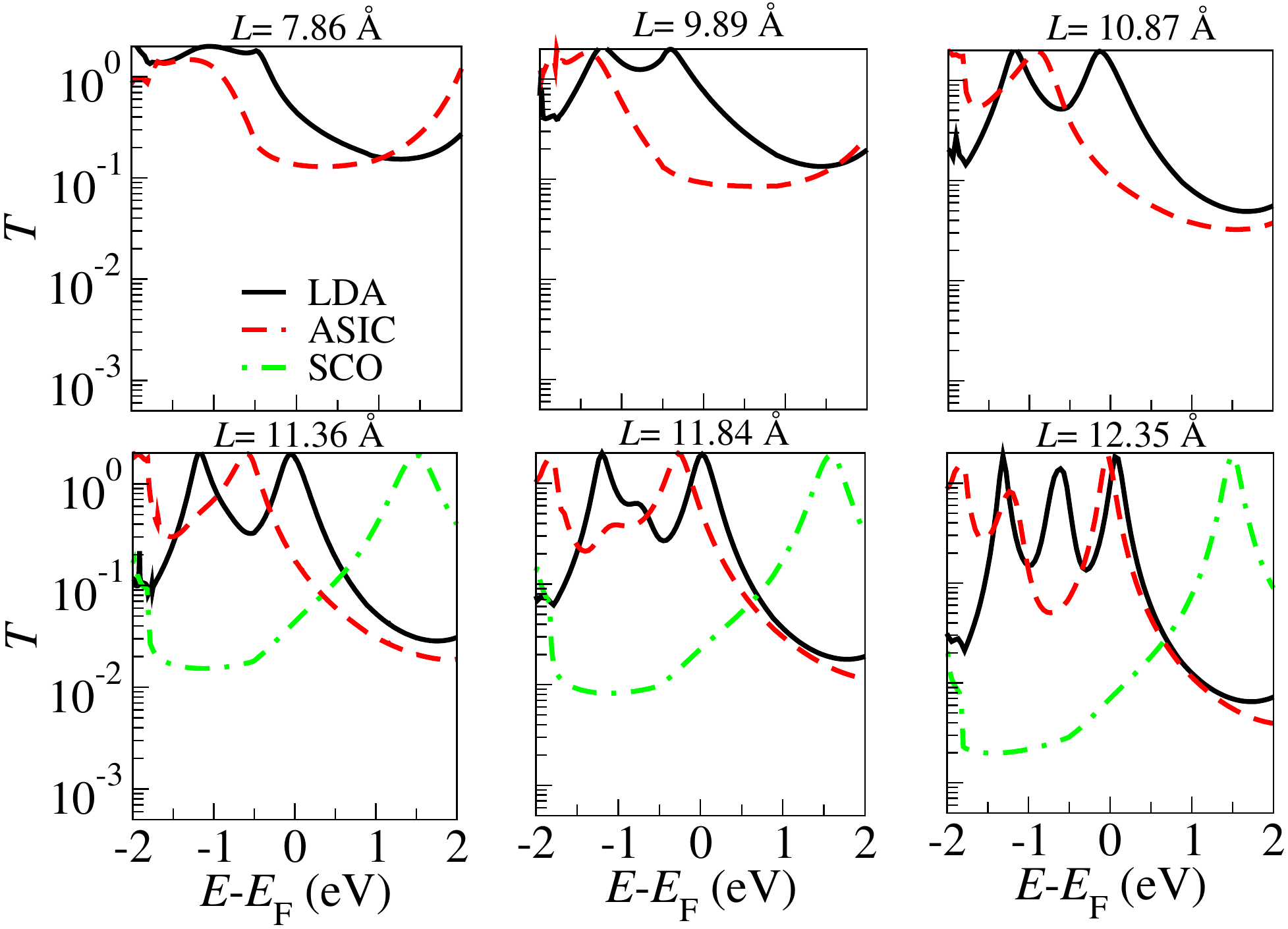}
\caption{Transmission coefficients as a function of energy for 
different electrode separation for the Au-BDT-Au junctions. Comparison between LDA, ASIC and LDA+SCO.}
\label{fig:bdtpics-trc-sco}
\end{figure}

\begin{figure}[!h]
\center
\includegraphics[width=0.46\textwidth]{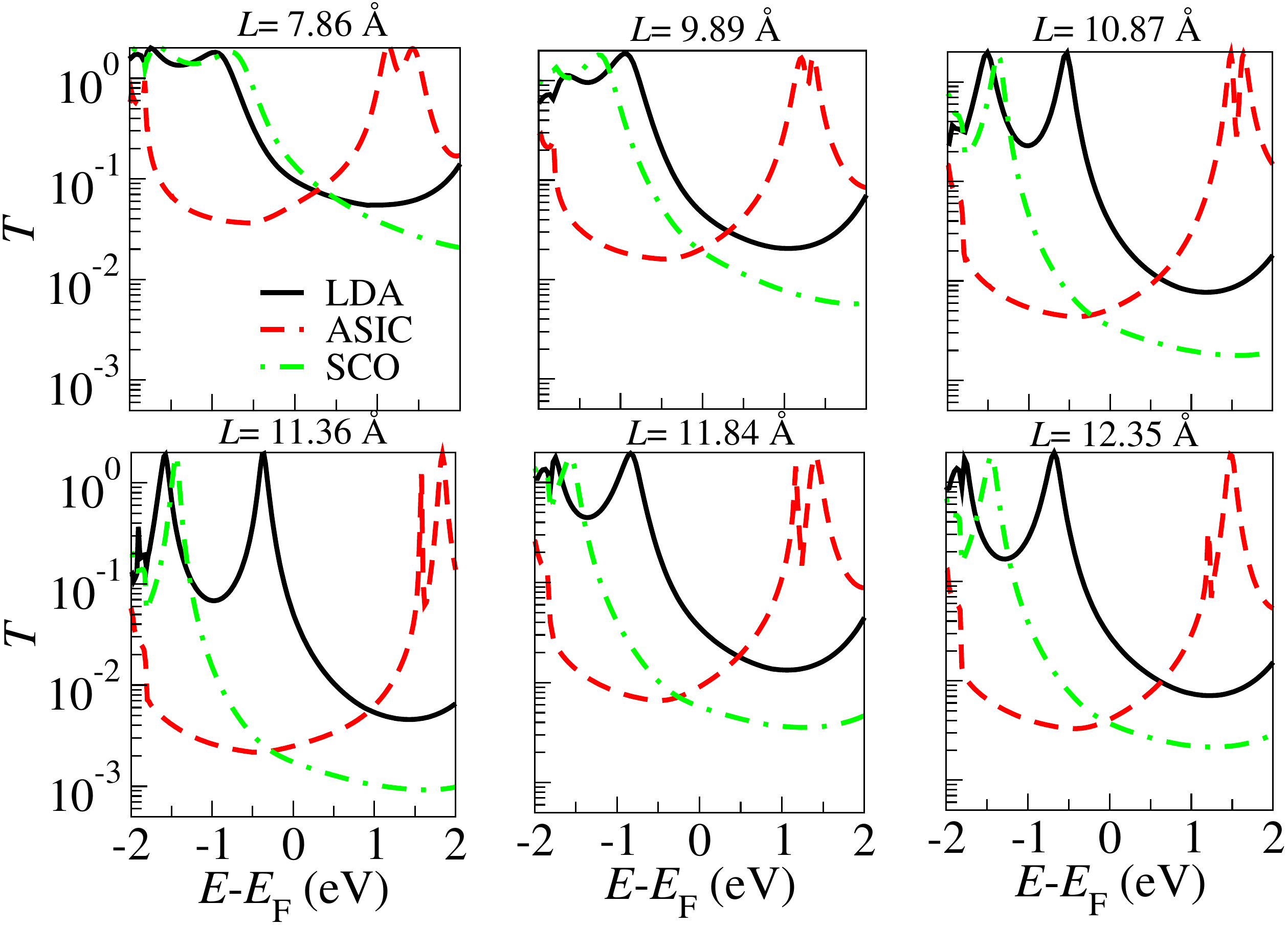}
\caption{Transmission coefficients as a function of energy for 
different electrode separation for the Au-BDT2H-Au junctions. Comparison between LDA, ASIC and LDA+SCO.}
\label{fig:bdt2hpics-trc-sco}
%\end{minipage}
\end{figure}

\begin{figure}[!h]
\center
\includegraphics[width=0.45\textwidth]{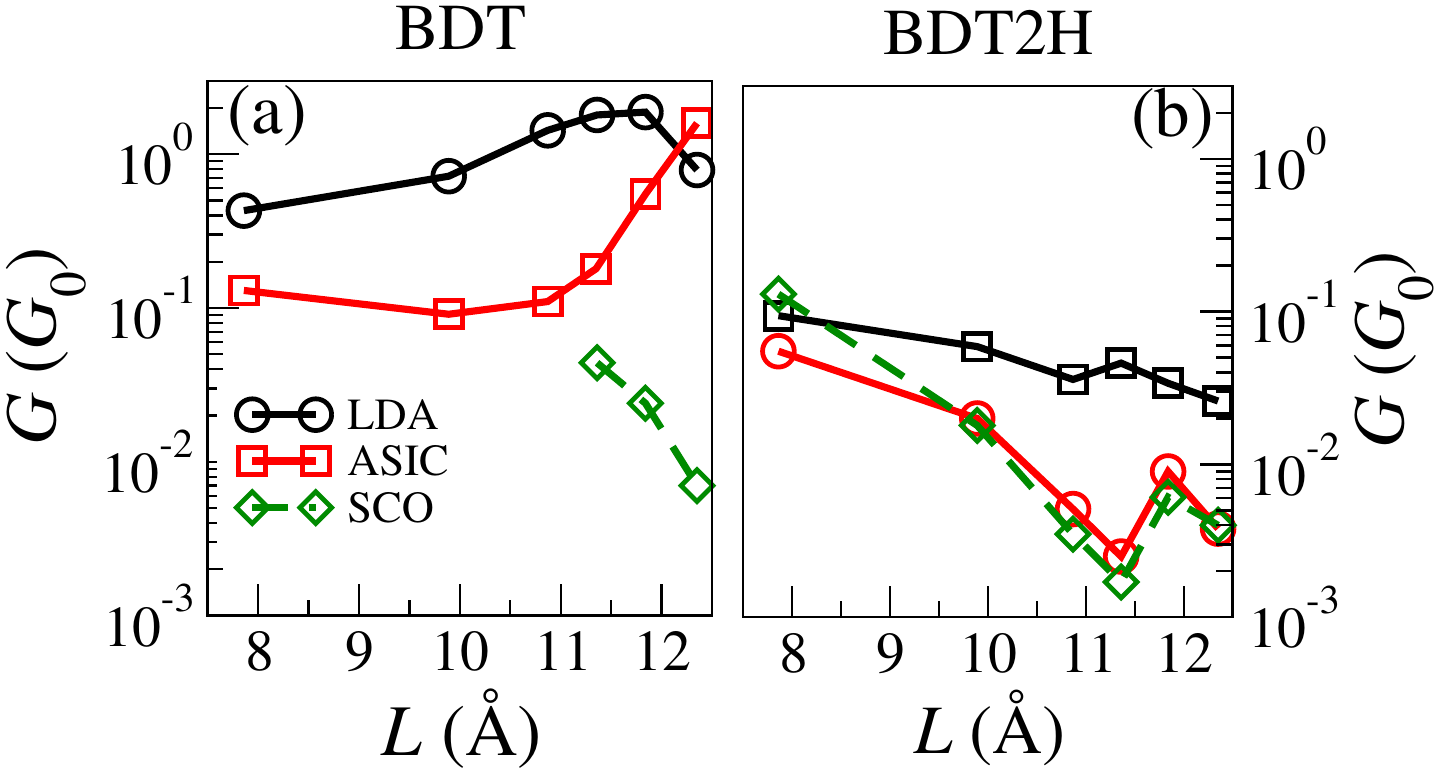}
\caption{Conductance as a function of $L$ for (a) Au-BDT-Au and (b) Au-BDT2H-Au molecular junctions.
Comparison between LDA, ASIC and scissor operator (SCO) results.}
\label{fig:conductance}
\end{figure}

Hereafter we present results for the transport properties as a function stretching,
of molecules attached to flat Au electrodes. Fig. \ref{fig:bdtpics-trc-sco} shows
the transmission coefficients for the Au-BDT-Au junctions corresponding to Figs. \ref{fig:bdtpics}(c)-(h),
while Fig. \ref{fig:bdt2hpics-trc-sco} shows the same for the Au-BDT2H-Au junctions of
Figs. \ref{fig:bdt2hpics}(c)-(h). We start by discussing the results for the Au-BDT-Au junctions.
In this case, the HOMO moves from lower energies at small $L$ towards $E_\mathrm{F}$ at larger $L$.
This results in an increase of $G$ under stretching [Fig. \ref{fig:conductance}(a)].
This is in agreement with previous theoretical works~\cite{Pontes2011a, Toher2008, Romaner2006, Sergueev2010}
for BDT attached to flat Au electrodes.
%trc_bdt_all_dist_lda_asic_sco
% 

Recently, using low-temperature MCBJs Bruot \textit{et al.}~\cite{Bruot2012} observed some
conductance traces where $G$ changed from 0.01$G_0$ to 0.1$G_0$ by increasing $L$.
The authors attributed this to the HOMO level moving up in energy towards the $E_\mathrm{F}$ of the electrodes.
However, most experimental 
results~\cite{Lortscher2007,Gonzalez2006,Kim2011, Taniguchi2010,Tian2006,Tian2010,Tsutsui2009,Tsutsui2006,Tsutsui2009a,
Quek2009, Arroyo2011,Baheti2008,Fatemi2011,Kiguchi2010,Reddy2007,Vazquez2012, Venkataraman2006,Wold2002,Tao2004,Xu2003}
show conductance traces with either approximately constant $G$ under stretching, or with decreasing $G$ for increasing $L$.\cite{Huang2006,Kim2011} 
In the calculations of French \textit{et al.}~\cite{French2013a} two types
of conductance traces are found: (i) large increase under stretching and (ii)
approximately constant values. The increase of $G$ is found only for junctions
that form monoatomic chains (MACs) of gold atoms connected to the BDT molecules.
MAC formation leads to an increase of the DOS at $E_\mathrm{F}$ in the contact Au atoms,
which adds to the increase of $G$ due to the HOMO shifting closer to $E_\mathrm{F}$ under stretching.

By applying the ASIC the absolute value of $G$ decreases by up to one order of magnitude
when compared to the LDA values, since the HOMO level is shifted to lower energies
(Fig. \ref{fig:bdtpics-trc-sco}). For small $L$ the ASIC $G$ vs. $L$ curve is approximately
constant, while for large $L$ the value of $G$ is found to increase for large $L$ [Fig. \ref{fig:conductance}(a)],
which is also due to the fact that the HOMO level ($\Psi_1$) is approaching $E_\mathrm{F}$
as the junction is stretched. Our CDFT results of the previous section show that
for $L\ge11.36$~\AA~the $\Psi_1$ state is expected to be located at least $\sim$1.5 eV
above $E_\mathrm{F}$. Thus we apply the SCO to shift the eigenvalue of $\Psi_1$ to
this energy, and calculate the transmission (green-dashed lines) and $G$ (for $L\ge11.36$~\AA~)
by using the calculated corrections presented in Table. \ref{bdt2h_energy_corrections}.
The corrected $G$ is smaller than the LDA results by up to two orders of magnitude and smaller than the ASIC by about
a factor of 10.

In contrast, for the Au-BDT2H-Au structures, $G$ decreases with increasing $L$ for all used XC functionals
(Fig. \ref{fig:bdt2hpics-trc-sco}). For LDA $G$ monotonically decreases from 0.1$G_0$ to 0.026$G_0$,
and using ASIC the values of $G$ further decrease by up to one order of magnitude. Applying the SCO
correction $E_\mathrm{LDA}^\mathrm{gap}$ increases, and consequently $G$ decreases by more than one
order of magnitude when compared to the LDA results, except for the shortest considered distance.
We note that although $G$ is similar for ASIC and SCO, $T$ at $E_\mathrm{F}$ is dominated by the
LUMO tail for ASIC (see Fig. \ref{fig:bdt2hpics-trc-sco}), while it is HOMO dominated for SCO.
The agreement between ASIC and SCO is mainly due to the fact that both put $E_\mathrm{F}$ in the gap, 
and the change of $G$ with stretching is mainly due to the change of the electronic
coupling to the electrodes.
The decreasing trend of $G$ vs. $L$ was observed by Ning \textit{et al.}~\cite{Ning2010}
where they considered the Au-BDT2H-Au junctions and the molecule is symmetrically
connected to an adatom structure. This is qualitatively in good agreement with the
experiments of Kim \textit{et al.}~\cite{Kim2011}, where by means of low-temperature MCBJ
technique, they reported values of $G$ ranging from $6.6\times10^{-4}$ to 0.5$G_0$ and
that high-conductance values are obtained when the molecular junction is compressed, i.e, $L$
decreases.
These are the key results of the present work since when combined with the results for
the formation energy of the hydrogenated junctions,
they indicate that the possibility of having thiol junctions can not be ruled out.
In fact, the thiol structures might be the ones
present in junctions where $G$ decreases with elongation.\cite{Huang2006,Kim2011} 

An important difference between the Au-BDT-Au and Au-BDT2H-Au junctions is the character of the charge carriers, i.e.,
whether it is hole-like or electron-like transport. For Au-BDT-Au, in the strong coupling limit where $L\le10.87$~\AA,
the charges tunnel through the tail of the HOMO-like level which leads to a hole-like transport (see top panel
of Fig. \ref{fig:bdtpics-trc-sco}). In the weak coupling limit, after considering the SCO, the charge
carriers tunnel through the tail of the LUMO-like level which leads to a electron-like transport,
as shown in the bottom panel of Fig. \ref{fig:bdtpics-trc-sco}. For Au-BDT2H-Au junctions,
the tunneling is always performed through the tail of the HOMO-like level (see Fig. \ref{fig:bdt2hpics-trc-sco})
and therefore the charge carriers are holes. This is an important information since, experimentally,
by means of thermoelectric transport measurements, it is possible to address which frontier molecular
level is the conducting level. It has been shown~\cite{Reddy2007} that for the systems discussed,
this level is the HOMO, which agrees with our findings for the Au-BDT2H-Au junctions and also
for the Au-BDT-Au junctions in the strong coupling limit. This leads to the conclusion that for
experiments where $G$ increases with stretching,~\cite{Bruot2012} the thiolate junction is present
and the explanation for this observed trend can be due to the formation of the MACs proposed by
French \textit{et al.}.\cite{French2013a} In contrast, the thiol structures might be the
ones present in experimental measurements showing the opposite trend.\cite{Huang2006,Kim2011}

\section{Conclusion}
We performed DFT calculations to study the adsorption process of methanethiol and BDT2H molecules on the Au(111) surface.
For all the structures studied we find that thiols are energetically more stable than their thiolate counterparts.
Moreover, we find a large activation barrier of about 1 eV for the the dissociation of the H atom from the thiol
groups adsorbed on Au(111). These results indicate that the non-dissociated structures
are likely to exist in experiments, and therefore can not be ruled out.

The energy level alignment between molecule and electrodes is one of the main factors that determine the conductance.
To overcome the limitations of using the LDA-DFT eigenvalues we apply a CDFT method, which is based on total energy
differences in the same way as $\Delta$SCF calculations, with the difference that it allows also the inclusion of the
non-local Coulomb interaction that leads to the renormalization of energy levels as the molecule is brought close
to a metal surface. We find a reduction of the BDT $E_\mathrm{QP}^\mathrm{gap}$ of 2.09 eV with respect to its
gas phase gap, when the molecule is brought closer to a single Au(111) surface. CDFT also allows us to obtain
the height of the image charge plane on Au(111), which we find to be at about 1~\AA~ above the gold surface.
While for the BDT2H molecules the coupling to the surface remains small at all distances,
for small molecule-surface separation the electronic coupling between BDT and Au becomes very strong,
and in this limit the use of the CDFT approach is not applicable. The strong coupling leads to a significant
electron transfer from the surface to the molecule, so that the molecular LUMO of isolated BDT becomes
increasingly occupied as the molecule-surface distance decreases. When we correct for the self-interaction
error in the LDA XC functional, the electron transfer is enhanced. At the equilibrium Au(111)-BDT bonding
distance we then find that the molecular LUMO of isolated BDT has become fully filled. On the other hand, for BDT2H, 
the filling of the molecular orbitals does not depend on the distance to Au.

By means of NEGF+DFT we have then calculated the transport properties of the junctions with different contact
geometries and compare the results obtained with LDA, ASIC and LDA+SCO functionals. For the thiol structures,
the LDA values for $G$ are about one order of magnitude smaller than their thiolate counterparts. ASIC leads to
values of $G$ in better agreement with experiments for the thiolate systems. However, ASIC also leads to a spurious
increase of $G$ for the thiol junctions due to the down-shift of the empty states towards $E_\mathrm{F}$,
an artifact avoided in the SCO approach. We find that Au-BDT-Au and Au-BDT2H-Au junctions show opposite trends concerning
the dependence of $G$ on the separation between flat Au electrodes; $G$ decreases with $L$ for the thiol
junctions, whereas the thiolates show the opposite trend. Since for Au-BDT2H-Au there is no significant charge
transfer between the electrodes and the molecule, we can apply the SCO approach to set the HOMO-LUMO gap to the one
obtained from CDFT calculations. In this way $G$ decreases by up to two orders of magnitude when compared
to the LDA values, and this brings the results in good quantitative agreement with the experimental data. Our results therefore
suggest that thiol junctions must be present in experiments where $G$ decreases with $L$. In contrast, thiolates structures 
are likely to be
present in experiments showing a increase of the conductance upon stretching.  
\section*{Acknowledgments}
Research reported in this publication was supported by the King Abdullah
University of Science and Technology (KAUST).
The Trinity College High-Performance Computer Center 
and the HPC cluster at Universidade de S\~ao Paulo provided the computational resources.

\end{document}